\documentclass[prd,amsmath,amssymb,preprintnumbers,showpacs]{revtex4}

\usepackage{latexsym}
\usepackage{bm}
\usepackage{amsthm,amscd}
\usepackage{graphicx}

\newcommand{\real}{\mathbb{R}}
\newcommand{\complex}{\mathbb{C}}
\newcommand{\naturals}{\mathbb{N}}
\newcommand{\A}{\mathcal{A}}
\newcommand{\B}{\mathcal{B}}

\newcommand{\D}{\mathcal{D}}

\newcommand{\F}{\mathcal{F}}

\newcommand{\J}{\mathcal{J}}

\newcommand{\M}{\mathcal{M}}
\newcommand{\N}{\mathcal{N}}
\newcommand{\R}{\mathcal{R}}
\renewcommand{\S}{\mathcal{S}}
\newcommand{\T}{\mathcal{T}}
\newcommand{\U}{\mathcal{U}}
\newcommand{\V}{\mathcal{V}}

\newcommand{\Z}{\mathcal{Z}}

\newcommand{\fT}{\mathfrak{T}}

\newcommand{\FF}{{\sf F}}
\newcommand{\HH}{{\sf H}}
\newcommand{\KK}{{\sf K}}

\newcommand{\QQ}{{\mathfrak{Q}}}

\newcommand{\bd}{{\bf d}}
\newcommand{\bdelta}{\bm{\delta}}
\newcommand{\bk}{{\boldsymbol{k}}}
\newcommand{\mass}{M}

\newcommand{\id}{\openone}

\newcommand{\WF}{{\rm WF}\,}
\newcommand{\supp}{{\rm supp}\,}
\newcommand{\ux}{\underline{x}}
\newcommand{\uk}{\underline{k}}
\newcommand{\stack}[2]{\substack{#1 \\ #2}}

\newtheorem{Definition}{Definition}[section]
\newtheorem{Theorem}[Definition]{Theorem}
\newtheorem{Proposition}[Definition]{Proposition}
\newtheorem{Lemma}[Definition]{Lemma}

\newcommand{\dip}[2]{\langle\langle #1\mid #2\rangle\rangle}

\begin{document}
\title{A Quantum Weak Energy Inequality for Spin-One Fields in Curved Spacetime}
\author{Christopher J. Fewster}
\email{cjf3@york.ac.uk}
\author{Michael J. Pfenning}
\email{mpj11@york.ac.uk}
\affiliation{Department of Mathematics, University of York, Heslington, York, YO10 5DD, UK}
\date{25 March 2003}
\begin{abstract}
Quantum weak energy inequalities (QWEI) provide state-independent lower bounds on 
averages of the renormalised energy density of a quantum field. We
derive QWEIs for the electromagnetic and massive spin-one fields in
globally hyperbolic spacetimes whose Cauchy surfaces are compact and
have trivial first homology group. These inequalities
provide lower bounds on weighted averages of the renormalized energy
density as ``measured'' along an arbitrary timelike trajectory,
and are valid for arbitrary Hadamard states of the spin-one fields. 
The QWEI bound takes a particularly simple form for averaging along static
trajectories in ultrastatic spacetimes; as specific examples we consider
Minkowski space [in which case the topological restrictions may be dispensed with]
and the static Einstein universe.

A significant part of the paper is devoted to the definition and
properties of Hadamard states of spin-one fields in curved spacetimes,
particularly with regard to their microlocal behaviour. 
\end{abstract}
\preprint{ESI 1295 (2003)}
\maketitle

\section{Introduction}

In common with all observed forms of classical matter, the spin-one
fields described by the Maxwell and Proca equations obey the weak energy
condition (WEC). That is, the stress-energy tensor $T_{ab}$ obeys
$T_{ab}v^a v^b\ge 0$ for all timelike vectors $v^a$. In classical
general relativity, energy conditions such as the WEC play a key
role in many important results, notably the singularity theorems of 
Penrose \cite{Penrose} and Hawking \cite{Hawking}. Moreover, since
any metric solves the Einstein equations for some choice
of stress-energy tensor, it is arguable that 
general relativity has limited predictive power in the absence of such
conditions. 

However, all classical energy conditions are violated by quantum
fields, as has been known for many years~\cite{Epstein}. Typically,
the energy density at a given spacetime point is unbounded from
below as a function of the state. Specific examples of negative
energy states are provided by highly squeezed states of light in
quantum optics, the Casimir vacuum state for a quantized field
between uncharged perfectly conducting parallel planar plates and
the Boulware vacuum state outside a black holes. Replacing the
classical stress-energy tensor on the right hand side of Einstein's
equation with its quantum expectation value, we must therefore
allow for negative energy sources, raising the possibility of
exotic phenomena within the realm of semiclassical gravity. For
example, negative energy can be used to maintain static traversable
wormholes~\cite{Morris88a,Morris88b}, create naked singularities
\cite{Hisc81,F&Ro92}, travel faster than light~\cite{Alcu94,Kras95},
and travel backward through time~\cite{Morris88b,Ever94}. In negative
energy models studied by Parker and Fulling it is even possible to
avoid the cosmological singularity~\cite{P&Fu73}, thus threatening
to overthrow the classical singularity theorems. One might also expect
that negative energy fluxes could lead to macroscopic violations
of the second law of thermodynamics~\cite{Ford78}.  

Of course, macroscopic violations of the second law are not observed in nature.
Motivated by these thermodynamic considerations, Ford~\cite{Ford78}
deduced that the negative fluxes and energy densities of quantum fields must
be subject to constraints, at least on average. These constraints,
known as {\em quantum inequalities}, or QI, provide state-independent lower bounds
on certain weighted averages of the stress-energy
tensor. They may therefore be regarded as the remnants of the classical energy
conditions after quantisation; thus, for example, the analogue of the 
WEC is sometimes referred to as a {\em quantum weak energy inequality} or
QWEI. Such bounds typically take the form
\begin{equation}
\int  \langle : T_{ab}(\gamma(\tau)) u^{a}(\tau) u^{b}(\tau):\rangle_\omega \, 
f(\tau) \, d\tau
\geq - \QQ(\gamma,f)\,,
\label{eq:QItyp}
\end{equation}
where $f\ge 0$ is the weight, or sampling function, $\gamma$ is a
smooth timelike curve, parametrised by proper time $\tau$ and with
four-velocity $u^a=(d\gamma/d\tau)^a$, and $:T_{ab}:$ denotes the stress-energy tensor,
normal ordered with respect to some reference state $\omega_0$. 
The significant point is that the bound $\QQ(\gamma,f)$ is independent
of the state $\omega$. The class of states $\omega$ for which the
bound holds must also be delineated---all bounds in the literature
require (at least) that $\omega$ and $\omega_0$ be Hadamard states,
although this can be weakened slightly. Of course the normal ordered
energy density differs from the renormalised energy density by
the (renormalised) energy density of the reference state. Accordingly, 
Eq.~(\ref{eq:QItyp}) may easily be converted into a bound on the renormalised
energy density. 

As we will shortly describe, various authors have established QWEI's
for the scalar and Dirac fields in different circumstances, leading up
to general results valid in curved spacetimes. 
The main aim of the present paper is to establish similar
QWEI's for the Maxwell and Proca fields in general globally hyperbolic
spacetimes. In order to do this, we have also made a detailed study
of the field theories concerned, particularly in relation to the class
of Hadamard states. Several new results obtained here may therefore be of
more general interest. 
 
As already mentioned, the earliest work on quantum inequalities is
due to Ford, who established a bound on negative energy fluxes for
scalar fields in 1991~\cite{Ford91}. The first QWEI was obtained
for the massless, minimally-coupled scalar field in Minkowski spacetime by
Ford and Roman~\cite{F&Ro95}, who established the inequality~(\ref{eq:QItyp})
for the case of an inertial worldline $\gamma$ and with $f$ taking the 
Lorentzian form $f(\tau)= \tau_0/(\pi(\tau^2+\tau_0^2))$, in which 
$\tau_0$ sets the characteristic timescale for the averaging. 
In four dimensions, the resulting bound was
\begin{equation}
\int  \langle : T_{ab}(\tau) u^{a}(\tau) u^{b}(\tau):\rangle_\omega \, 
\frac{\tau_0}{\pi(\tau^2+\tau_0^2)}\, d\tau
\geq - \frac{3}{32\pi^2\tau_0^4}\,.
\label{eq:QI}
\end{equation}
These results were then extended by Ford and Roman to the massive
scalar field in Minkowski space~\cite{F&Ro95,F&Ro97} and were further
extended by Pfenning and Ford to minimally-coupled scalar fields of
arbitrary mass in ultrastatic curved spacetimes \cite{Pfen97a,
Pfen98a,Pfen98b}. In these generalisations, it was found that the dominant term of the
QWEI has a form similar to Eq.~(\ref{eq:QI}), but with subdominant correction terms due 
to the curvature and the mass of the field. Pfenning and Ford 
also showed that one could express
the bound on the right hand side of the QWEI in terms of derivatives of the
Euclidean Green's function for the spacetime and developed a short
sampling time approximation to the QWEI which could be used in spacetimes
where it would be too hard to calculate the exact QWEI bound.

The restriction to the Lorentzian weight
$\tau_0/(\pi(\tau^2+\tau_0^2))$ was removed by various authors. 
By making use of the conformal properties of
field theories in two dimensions, Flanagan \cite{Flan97,Flan02} has
derived optimal quantum inequalities for the massless scalar
field for arbitrary smooth positive sampling functions, and Vollick has done
the same for the Dirac field \cite{Voll00}. QWEI's for the minimally coupled
scalar field in static curved spacetimes of any dimension for an arbitrary
sampling function were established by Fewster in work with
Eveson~\cite{Fe&E98} and Teo~\cite{Fe&T99}. 

More recently, techniques drawn from microlocal analysis have been used
to considerably generalise previous QWEI's and to put them on a
mathematically rigorous footing. Fewster~\cite{Fews00} used these techniques
to derive a QWEI for minimally coupled scalar fields in general globally hyperbolic
spacetimes (the most general class on which the Klein--Gordon equation is
well-posed). In this case, averaging takes place along an arbitrary timelike
worldline, using any weight of the form $f(t)=g(t)^2$ for $g$ smooth,
real-valued and compactly supported. Subsequently, Fewster and Verch
established similar results for the Dirac and Majorana fields in
four-dimensional globally hyperbolic spacetimes~\cite{Fe&V02}.
Averaging over spacetime volumes has been considered by
Helfer~\cite{HelferII}, also in great generality. 

Previous work on QWEIs for spin-one fields has focussed on the
electromagnetic field, beginning with the work of Ford and
Roman~\cite{F&Ro97}, who derived a QWEI for the case of Lorentzian
sampling along inertial trajectories in Minkowski 
space. More recently, Pfenning~\cite{Pfen02} has derived a QWEI for the
electromagnetic field in static curved spacetimes with arbitrary positive
weight sampling functions by using the techniques developed for the 
scalar field in~\cite{Fe&E98,Fe&T99}. Using similar techniques, 
Marecki~\cite{Marecki} has derived bounds on the fluctuations of the
electric field strength, which are of interest in quantum optics. 

In this paper we will adapt the methods of~\cite{Fews00} to the Maxwell
and Proca fields. This depends crucially on the fact, first discovered
by Radzikowski~\cite{Radz96} for scalar fields, that the class of Hadamard states
may be characterised in terms of a wave-front set condition on the two-point
function. Similar reformulations are known for the Dirac 
field~\cite{Kratzert,Hollands,Sah&Ver}
but there is as yet no full treatment for the Maxwell and
Proca fields in the literature. (See, however,~\cite{B&OT86,AllOtt92} for
[non-microlocal] discussions of Hadamard states for electromagnetism
using Faddeev--Popov ghosts.) Our treatment of this issue has been influenced to
some extent by the forthcoming work of Junker and Lled\'o~\cite{JunLle},
although it has been conducted largely independently, leading to some technical
differences with their approach~\footnote{We also observe that their
work is directed towards a construction of Hadamard states in general
globally hyperbolic spacetimes which avoids the use of deformation
arguments.}.  

The paper is structured as follows. The Maxwell and Proca fields are
most elegantly described using differential forms; accordingly, 
we begin in Section~\ref{sec:prelims} with a description of our
conventions for differential forms and other geometric objects which
will appear in this paper. In particular, we delineate the class of
globally hyperbolic spacetimes to be considered; for
technical reasons it is convenient to assume that their Cauchy surfaces
are compact and have trivial first homology group. 
This is followed by a brief introduction to microlocal analysis leading
to the definition of the wave-front set for $p$-form distributions. 

In Sec.~\ref{sect:quantisation} we describe the quantisation of the
Maxwell and Proca fields in globally hyperbolic spacetimes. We adopt an
algebraic approach, giving a direct construction of algebras of
observables equivalent to those obtained by Dimock~\cite{Dimock} and
Furlani~\cite{Furl99}. We also define the notion of a Hadamard state for
these fields, as a state whose two-point function~\footnote{Some
care is needed over the definition of the two-point function in the Maxwell
case (see Sec.~\ref{sect:quantisation}). In particular, it is not a distribution.}
is related in a certain way to a one-form {\em Klein--Gordon} bisolution of
Hadamard form. Such bisolutions and their microlocal properties 
have been discussed in detail by Sahlmann and Verch~\cite{Sah&Ver}; we
may therefore read off the microlocal properties of the Maxwell and
Proca two-point functions in Hadamard states. This permits us to apply the
methods of~\cite{Fews00} to obtain quantum inequalities for these fields
in Sec.~\ref{sect:QWEI}. Before this, in Sec.~\ref{sect:ultra}, we specialise to the
class of ultrastatic spacetimes in order to gain further insight into
the abstract definitions of Sec.~\ref{sect:quantisation}. In particular, 
we prove the existence of Hadamard states in these spacetimes and use 
deformation arguments~\cite{FNW81} to deduce from this the existence of
Hadamard states in general globally hyperbolic spacetimes obeying our
topological restrictions. We also compare our approach with other
quantisation schemes, including the Gupta--Bleuler method. As mentioned
above, we expect that some of the results obtained here to be of wider
interest. 

In Sec.~\ref{sect:examples} we investigate our QWEI in Minkowski space,
and in general ultrastatic spacetimes (modulo the usual topological
conditions). Simple formulae are obtained for the QWEI bound, which are
readily compared with those previously obtained for the scalar field.
In Minkowski space, the QWEI bound is weaker by a factor of exactly two
for the Maxwell field (as already noted in~\cite{Pfen02}) and by a
factor of three for the Proca field. This is not very surprising and
simply reflects the number of spin degrees of freedom. In curved
spacetime, however, the spin-one and scalar
QWEIs cannot be related in this fashion. To emphasise this, we
explicitly determine the QWEI bound in the Einstein static universe,
providing a concrete example of our ultrastatic QWEI. 
Two appendices contain the proofs of some
technical results required in the body of the paper.

\section{Preliminaries}\label{sec:prelims}

Units where $\hbar = c =1$ are used throughout. The notation
$C_0^\infty(\real^n)$ denotes the space of smooth, compactly
supported~\footnote{The {\em support} of a function is the
closure of the set of points on which it is nonzero.},
complex-valued functions on $\real^n$.

\subsection{Geometry and forms}
\label{sect:forms}

Spin-one fields on globally hyperbolic spacetimes are most elegantly
formulated in terms of differential forms. We will follow
the conventions of~\cite{AMR88}, which we now briefly summarise for the
benefit of the reader. 

Suppose $\N$ is a smooth $n$-dimensional manifold
which is connected, boundaryless, orientable, Hausdorff, paracompact,
and equipped with a smooth metric of index
$s$~\footnote{The index is the number of spacelike (i.e., negative
norm-squared) basis vectors in any $g$-orthonormal frame.}.
We denote the space of smooth, complex-valued $p$-forms on $\N$ by
$\Omega^p(\N)$; the subspace of compactly supported $p$-forms will be
written $\Omega_0^p(\N)$. Each $p$-form may be regarded as an 
antisymmetric covariant $p$-tensor field and we will occasionally use index
notation accordingly. Thus the exterior product
$\alpha\wedge\beta\in\Omega^{p+q}(\N)$ of
$\alpha\in\Omega^p(\N)$ and $\beta\in\Omega^q(\N)$
is given by
\begin{equation}
(\alpha\wedge \beta)_{a_1\cdots a_{p+q}} = \frac{(p+q)!}{p!q!}
\alpha_{[a_1\cdots
a_p}\beta_{a_{p+1}\cdots a_{p+q}]}\,,
\end{equation}
and the exterior derivative $\bd:\Omega^p(\N)\to\Omega^{p+1}(\N)$ is
defined by
\begin{equation}
(\bd\alpha)_{a_1\cdots a_{p+1}}=(p+1)\nabla_{[a_1}\alpha_{a_2\cdots
a_{p+1}]}\,,
\end{equation}
where the square brackets denote antisymmetrisation and $\nabla_a$ is
any connection on $\N$ ($\bd$ is independent of the choice of
connection). The Hodge $*$-operator is defined uniquely as
the map $*:\Omega^p(\N)\to \Omega^{n-p}(\N)$ such that
\begin{equation}
\alpha\wedge *\beta = \frac{1}{p!} \alpha_{a_1\ldots a_p}
\beta^{a_1\cdots a_p} d{\rm vol}_g\,,
\end{equation}
where $d{\rm vol}_g$ is the positive volume $n$-form associated with the
metric $g$. In particular, $(*)^2=(-1)^{p(n-p)+s}$ on $p$-forms. By combining the
Hodge $*$ and exterior derivative, we may define the
coderivative $\bdelta:\Omega^{p}(\N)\to\Omega^{p-1}(\N)$  
(with the convention that $\bdelta$ annihilates all zero-forms) by
$\bdelta = (-1)^{n(p-1)+s+1} *\bd *$, which reduces to $\bdelta= *\bd *$
in a four-dimensional Lorentzian spacetime.

The operations introduced above allow us to define a symmetric pairing
$\langle\cdot,\cdot\rangle$ of $p$-forms under integration: we set
\begin{equation}
\langle \U ,\V \rangle \equiv \int_\N \U \wedge *\V
\end{equation} 
for any $\U,\V\in\Omega^p(\N)$ for which the integral exists. Since
$\bd(\U\wedge*\V) = \bd\U\wedge*\V -\U\wedge*\delta\V$, Stokes' theorem gives
\begin{equation}
\langle \bdelta\U,\V\rangle = \langle \U,\bd\V\rangle
\label{eq:d_delta_duality}
\end{equation}
for smooth $(p+1)$- and $p$-forms $\U$ and $\V$ whose supports
have compact intersection. In this sense the operators 
$\bd$ and $\bdelta$ are dual. 

The Laplace-Beltrami operator is 
defined as $-(\bd\bdelta + \bdelta\bd)$, i.e.,
it is equal to {\em minus} the Laplace--de~Rahm operator 
$\bd\bdelta+\bdelta\bd$. Where the manifold $\N$ is a
Lorentzian spacetime, with signature $+-\cdots-$, the Laplace-Beltrami operator
is also known as the D'Alembertian and will be denoted by $\Box$. We wish
to point out that Dimock~\cite{Dimock} uses $\Box$ to denote the
Laplace--de~Rahm operator; our usage is determined by the convention
that $\Box$ should have principal part $g^{\mu\nu}
\partial_\mu \partial_\nu$, in accordance 
with typical usage in general relativity.

Finally, the spaces $\Omega_0^p(\N)$ may be given locally
convex topologies~\footnote{The topology is 
defined so that a sequence $\U_n\to 0$ in $\Omega_0^p(\N)$ if and only
if $*(\V\wedge *\U_n)\to 0$ in $\D(\N)$ for every $\V\in\Omega^p(\N)$.}
and the corresponding topological duals
$\Omega_0^p(\N)'$ will be called the spaces of $p$-form distributions on $\N$. 
(The pairing $\langle\cdot,\cdot\rangle$ provides a natural embedding of
$\Omega^p(\N)$ in $\Omega_0^p(\N)'$.) In the case $p=0$, we will also use
$\D(\N)$ and $\D'(\N)$ for $\Omega_0^0(\N)$ and $\Omega_0^0(\N)'$
respectively. The exterior derivative and coderivative are defined on
these spaces by
\begin{equation}
(\bd\U)(f) = \U(\bdelta f) \qquad
(\U\in\Omega_0^p(\N)',~f\in\Omega_0^{p+1}(\N))
\end{equation}
and
\begin{equation}
(\bdelta\U)(f) = \U(\bd f) \qquad
(\U\in\Omega_0^{p}(\N)',~f\in\Omega_0^{p-1}(\N))
\end{equation}
which extend the definitions given for smooth forms by virtue of the
embeddings defined above and the calculation~(\ref{eq:d_delta_duality}).
In a similar way, the Hodge $*$-operator may also be extended to a map
from $p$-form distributions to $(n-p)$-form distributions by
\begin{equation}
(*\U)(f) =(-1)^{p(n-p)}\U(*f) \qquad (\U\in\Omega_0^p(\N)',~f\in\Omega_0^{n-p}(\N))\,;
\end{equation}
and it is easily checked that the formula $\bdelta=(-1)^{n(p-1)+s+1} *\bd *$ remains true for
distributions.

\subsection{Microlocal analysis and the wave-front set}

Our proof of the quantum weak energy inequality turns on the detailed
singularity properties of various distributions related to the two-point
functions of quantum fields in Hadamard states. The information required
is encoded in the {\em wave-front set} of these distributions, which is
defined as follows. (See Ref.~\cite{Hormander_1} for a full presentation.)

We will define the Fourier transform $\widehat{u}$ of 
$u \in C_0^\infty(\real^n)$ using the nonstandard convention
\begin{equation}
\widehat{u}(k) = \int d^ny\ u(y) e^{ik\cdot y},
\end{equation}
which conforms to the conventions used e.g., in \cite{Fews00}.
The
Fourier transform can be extended to scalar distributions of compact
support by writing $\widehat{u}(k)=u(f_k)$ where $f_k=e^{ik\cdot y}$.
Given a cone $V\subset(\real^n\backslash\{0\})$, we will say that
$\widehat{u}(k)$ is of rapid decrease in $V$ if for each $N\in\naturals$
there exists a real constant $C_N$ such that
\begin{equation}
\left|\widehat{u}(k)\right| \leq \frac{C_N}{(1+|k|)^N}\qquad
\forall\ k\in V,
\end{equation}
where $|k|$ denotes the Euclidean norm of $k$. 

Smooth compactly supported functions have Fourier transforms which
decay faster than any inverse power in the whole of $\real^n$, but
the same is not true for arbitrary distributions $u$ of compact
support. A well known example is the Dirac
$\delta$-function, whose Fourier transform does not decay at infinity in any direction.
We define the set of singular directions $\Sigma(u)$ to be the set
of all $k \in \real^n \backslash\{0\}$ having no conical neighbourhood
$V$ in which $\widehat{u}$ is of rapid decrease. 

More detailed information about the singularities of $u$ can be gained
by localizing the singular directions. In particular, the set of 
{\it singular directions} of $u$ at a point $x$ is defined by
\begin{equation}
\Sigma_x(u) = \bigcap_{\chi}\Sigma(\chi u)\,,
\end{equation} 
where the intersection is taken
over all smooth compactly supported test functions $\chi\in C^\infty_0(\real^n)$
with $\chi(x)\neq 0$. The {\it wave-front set} $\WF(u)$ of $u$ is then
defined by
\begin{equation}
\WF(u)=\{ (x,k)\in \real^n\times(\real^n\backslash\{0\})\ |\ k\in\Sigma_x(u) \}. 
\end{equation}

The wave-front set can be extended in a natural way to distributions on
manifolds. Let $(\N,g)$ be a smooth $n$-dimensional manifold of the
type discussed above. Each
distribution $u$ in $\D'(\N)$ has a representative
$u_\kappa\in\D'(\real^n)$ in each
coordinate chart $(U,\kappa)$ (with corresponding coordinates denoted 
$x^\mu=\kappa(x)^\mu$) defined so that
\begin{equation}
u_\kappa(f\sqrt{-|g|}) = u(f\circ \kappa)
\label{eq:ukappa}
\end{equation}
holds for every smooth function $f$ compactly supported in
$\kappa(U)\subset \real^n$, where $|g|$ is the determinant of
$g_{\mu\nu}$. The wave-front set $\WF(u)$ is now a subset of the
cotangent bundle $T^*\N$ with the property that 
$(x,k)\in\WF(u)$ if and only if there is a chart $(U,\kappa)$ about $x$
so that $(\ux,\uk)\in\WF(u_\kappa)$, where 
$(\ux,\uk)$ are the coordinates of $(x,k)$ and $u_\kappa$ is defined as
in~(\ref{eq:ukappa}). 
In fact, it may be shown (see Theorem~8.2.4 and the following discussion
in Ref.~\cite{Hormander_1}) that the restriction of $\WF(u)$ to
$U$ is given by 
\begin{eqnarray}
\WF(u)\cap T^*U &=& \kappa^*\WF(u_\kappa) \nonumber\\
&:=& \{
\kappa^*(\ux,\uk): (\ux,\uk)\in\WF(u_\kappa)\}\,,
\end{eqnarray}
where the pull-back $\kappa^*$ relates $(x,k)\in T^*\N$ to its
coordinates by $\kappa^*(\ux,\uk)=(x,k)$.

There is a natural extension of the wavefront set to $p$-form distributions.
Let $v_i$ $(i=0,\ldots,n-1)$ be a global orthonormal $n$-bein on $\N$
(i.e., the vectors $v_i$ obey $g_{ab}v_i^a v_j^b=\eta^{ij}$ and thus
$\eta^{ij}v_i^a v_j^b=g^{ab}$, where $\eta^{ij}$ is diagonal with
$s$ entries equal to $-1$ and the rest equal to $+1$)
and let $V^i$ be the dual basis of one-forms: $V^i_a =
\eta^{ij}g_{ab}v^b_j$. For each
$\U\in\Omega_0^p(\N)'$, we may define the component distributions
$\U_{i_1\cdots i_p}\in \D'(\N)$ by
\begin{equation}
\U_{i_1\cdots i_p}(f) =\eta_{i_1 j_1}\cdots \eta_{i_p j_p}
\U(f V^{j_1}\wedge \cdots \wedge V^{j_p})
\end{equation}
for $f\in\D(\N)$. Then we define the wave-front
set of $\U$ to be
\begin{equation}
\WF(\U) = \bigcup_{i_1,\dots,i_p} \WF(\U_{i_1\dots i_p})\,,
\end{equation}
which may be shown to be independent of the particular choice of $n$-bein $v_i$.

\section{Quantisation of the Maxwell and Proca fields}
\label{sect:quantisation}

\subsection{Classical theory}

We will consider the Maxwell and Proca fields propagating on 
four-dimensional Lorentzian globally hyperbolic spacetimes.
Each such spacetime is a pair $(\M,g)$ consisting of a four-dimensional,
smooth, real manifold $\M$, with the topological properties listed at
the start of Sec.~\ref{sect:forms}, together with a smooth
Lorentzian metric $g$ with signature $(+,-,-,-)$. Global hyperbolicity,
which will ensure well-posedness of our field equations,
requires that $(\M,g)$ be time-orientable and that $\M$ contain
a Cauchy surface $\Sigma$, that is, a smooth spacelike hypersurface
intersected precisely once by every inextendible causal curve in $\M$.
In fact, one may show~\cite{Dieck88} that $\M$ is diffeomorphic to
$\real\times\Sigma$. We will assume for the most part that
$\Sigma$ [also referred to as the
spatial section] is compact and that $\Sigma$ has trivial first singular homology group
with real coefficients, $H_1(\Sigma)$. 
This is equivalent to the triviality of the
compact support de~Rahm cohomology group $H_c^3(\M)$ \footnote{Since
$\Sigma$ is compact, it admits a finite good cover [Theorem 5.1
in~\cite{BottTu}] which determines a finite good cover of
$\M=\real\times\Sigma$. Accordingly, we may use 
Poincar\'e duality [Theorem 5.4 in~\cite{BottTu}] 
to prove that $H_c^3(\M)$ is isomorphic to $H^1(\M)$ and hence 
to $H^1(\Sigma)$ [see the remarks following Corollary 5.1.1
in~\cite{BottTu}]. This is isomorphic to $H_1(\Sigma)$ by de~Rahm's
theorem~\cite{Warner71}.}; note that this condition
excludes, for example, the 3-torus $T^3$ as a spatial section. 
The compactness assumption is made for convenience only; 
triviality of $H_1(\Sigma)$ is inessential
for the Proca field, but appears to be required in order to establish
some of our results for the Maxwell field.

The classical uncharged spin-$1$ field of mass $\mass>0$ is a real
one-form field $\A\in\Omega^1(\M)$ obeying the Proca equation
\begin{equation}
\left(-\bdelta \bd + \mass^2\right) \A= 0\,.
\label{eq:proca_wave_eq}
\end{equation} 
Applying the coderivative $\bdelta$ we see that $\bdelta\A=0$ and so any
solution to Eq.~(\ref{eq:proca_wave_eq}) also satisfies the one-form
Klein--Gordon equation
\begin{equation}
\left(\Box + \mass^2\right) \A= 0\,.
\label{eq:hyper_wave_eq}
\end{equation}
Conversely, any solution to~(\ref{eq:hyper_wave_eq}) satisfying the
constraint
\begin{equation}
\bdelta\A = 0
\label{eq:constraint}
\end{equation}
solves the Proca equation. The advantage of the
system~(\ref{eq:hyper_wave_eq},\ref{eq:constraint}) is
that~(\ref{eq:hyper_wave_eq}) has the hyperbolic principal part 
$g^{\alpha\beta}\partial_\alpha \partial_\beta$
in local coordinates and therefore admits unique fundamental 
solutions $E^\pm_\mass :\Omega^1_0(\M)\rightarrow\Omega^1(\M)$ respectively
\cite{Choq67,Furl99}, such that
\begin{equation}
\left(\Box + \mass^2\right) E^\pm_\mass = E^\pm_\mass \left(\Box + \mass^2\right) = \id
\end{equation} 
and $\supp(E^\pm_\mass \J)\subset J^\pm(\supp \J)$
for $\J\in\Omega^1_0(\M)$. Here, $J^\pm(S)$, the causal future($+$)/past($-$)
of $S$, is defined to be the set of points in $(\M,g)$ that can be reached from the
set $S\subset\M$ by a future($+$)/past($-$) directed causal curve.
The operators $E^\pm_\mass $ extend to $\J$ with
$\supp \J$ compact to the past/future and for such $\J$, $\A^\pm
=E^\pm_\mass \J$ is the unique solution of $\left(\Box + \mass^2\right) A^\pm = \J$
with $\supp\A$ compact to the past/future. In addition, we
introduce the advanced-minus-retarded
bisolution~\footnote{Note that Dimock~\cite{Dimock} uses $E$ for the
retarded-minus-advanced bisolution.}
\begin{equation}
E_\mass =E^-_\mass - E^+_\mass\,,
\end{equation}
which satisfies the homogenous Klein--Gordon equation
\begin{equation}
\left(\Box + \mass^2\right) E_\mass = E_\mass \left(\Box + \mass^2\right) =0\,.
\end{equation}
We also note that, since $\bd$ and $\bdelta$ commute with the
Klein--Gordon operator $\Box+\mass^2$ (or, more
precisely, intertwine its action on zero- and one-forms)
these operators also intertwine the action of $E_\mass^\pm$ on zero- and
one-forms. 

The fundamental solutions for the one-form Klein--Gordon equation allow us
to solve the inhomogeneous Proca equation, 
\begin{equation}
\left(-\bdelta \bd + \mass^2\right) \A= \J \,,
\label{eq:inhom_proca}
\end{equation} 
with advanced ($-$) or retarded ($+$)
boundary conditions. Assuming the existence of a solution and applying the
coderivative, we find $\bdelta\A = \mass^{-2}\bdelta\J$. This allows us to
rewrite~(\ref{eq:inhom_proca}) as
\begin{equation}
\left(\Box + \mass^2\right)\A = \left(-\bdelta \bd + \mass^2\right) \A -
\bd\bdelta\A = \J - \mass^{-2}\bd\bdelta\J\,,
\end{equation}
to which $\A^\pm = E_\mass^{\pm}\left(\J-\mass^{-2}\bd\bdelta\J\right)$ are
the unique solutions with support in $J^\pm(\supp\J)$. Using the
property $\bdelta E_\mass^\pm=E_\mass^\pm\bdelta$, these may be shown to 
be the required solutions to~(\ref{eq:inhom_proca}). We write
\begin{equation}
\Delta_\mass^\pm = E_\mass^\pm\left(\id - \mass^{-2}\bd\bdelta\right)
\end{equation}
for the corresponding solution operators and define
\begin{equation}
\Delta_\mass = \Delta_\mass^- - \Delta_\mass^+\,,
\label{eq:DeltaM}
\end{equation}
which will later appear in the commutation relations for the quantised Proca field.
We will also use the notation $\Delta_M$ and $E_M$ for the
bidistributions defined by $\Delta_M(f,g)=\langle f,\Delta_M g\rangle$
and $E_M(f,g)=\langle f,E_M g\rangle$ ($f,g\in\Omega_0^1(\M)$). 

Turning to electromagnetism, the theory is described by a one-form potential
$\A\in\Omega^1(\M)$ obeying
\begin{equation}
\bdelta\bd\A=0\,,\label{eq:EM_wave_eq}
\end{equation}
which entails that the field strength $\F=\bd\A$ 
obeys Maxwell's equations, 
\begin{equation}
\bd \F = 0 \quad\mbox{and}\quad \bdelta \F = 0\,.\label{eq:Maxwell}
\end{equation}
Two potentials $\A$ and $\A'$ are gauge equivalent, denoted by $\A\sim\A'$, if $ \A =
\A'+\bd\chi$ for some $\chi\in\Omega^0(\M)$. Since gauge equivalent
potentials lead to the same field strength, it is really the gauge equivalence
classes $[\A]$ of solutions which are of physical significance rather
than the potentials themselves. We can partially fix the gauge freedom
by passing to the Lorentz gauge $\bdelta\A=0$, in which the Maxwell
equations are expressed by the $\mass=0$ case of the
system~(\ref{eq:hyper_wave_eq},\ref{eq:constraint}). Just as with the
Proca equation, the
fundamental solutions to~(\ref{eq:hyper_wave_eq}) permit us to solve the
inhomogeneous Maxwell equation 
\begin{equation}
-\bdelta\bd \A = \J\,,
\label{eq:inhom_Maxwell}
\end{equation}
where, for consistency, the source $\J\in\Omega_0^1(\M)$ is required to
obey the current conservation equation $\bdelta\J=0$.
Defining $\A^\pm=E_0^\pm\J$,
we note that $\bdelta\A^\pm=\bdelta E_0^\pm\J=E_0^\pm \bdelta\J
= 0$ and therefore deduce that $\A^\pm$ is the unique Lorentz gauge
solution to~(\ref{eq:inhom_Maxwell}) with support in $J^\pm(\supp\J)$.


\subsection{Quantisation: Algebras of Observables}

The canonical quantisation of the Maxwell and Proca fields in
globally hyperbolic spacetimes was accomplished by Dimock~\cite{Dimock}
and Furlani~\cite{Furl99} respectively (see also~\cite{Stro00}). 
Here, we give a direct construction
of suitable algebras of (polynomials in) smeared fields for these
theories, isomorphic to those emerging from the constructions of \cite{Dimock,
Furl99}. 
We begin with the more straightforward
Proca case, for which the algebra of observables will be denoted
$\mathfrak{A}_\mass (\M,g)$. This algebra is constructed by first using
the set of smooth compactly 
supported complex valued one-form test functions $\Omega^1_0(\M)$ to 
label a set of abstract objects $\{ \A(f) | f\in\Omega^1_0(\M)\}$
which generate a free unital $*$-algebra over $\complex$. These objects
are interpreted as smeared one-form fields
\begin{equation}
\A(f)\;\mbox{`}{=}\mbox{'}\;\langle\A, f\rangle\,.
\end{equation}
The algebra 
$\mathfrak{A}_\mass(\M,g)$ is defined to be the quotient of $\mathfrak{A}
_\mass$
by the relations:
\begin{enumerate}
\item[P1.] Linearity: $\A \left(\alpha f_1 + \beta f_2 \right) = \alpha\A(f_1)
+\beta\A(f_2)$ for all $\alpha,\beta\in\complex$ and $f_i\in\Omega^1_0(\M)$;
\item[P2.] Hermiticity: $\A(f)^* = \A(\overline{f})$ for all $f \in \Omega^1_0(\M)$;
\item[P3.] Field Equations: $\A \left( [\bdelta \bd -\mass^2] f \right)
= 0$ for all $f \in \Omega^1_0(\M)$;
\item[P4.] CCRs: $\left[\A(f_1),\A(f_2)\right] = -i\Delta_\mass (f_1, f_2)\id$
for all $f_i\in\Omega^1_0(\M)$. 
\end{enumerate}
Here, $\Delta_\mass$ is the propagator defined in~(\ref{eq:DeltaM}). 

For electromagnetism, quantisation is complicated by gauge freedom.
Even in Minkowski space this presents serious problems: as shown by 
Strocchi~\cite{StrocchiI,StrocchiIII} in the Wightman
axiomatic approach, the vector potential cannot exist as an operator-valued 
distribution if it is to transform correctly under the Lorentz 
group or even display commutativity at spacelike separations in a 
weak sense. Dimock~\cite{Dimock} circumvented such problems 
by constructing smeared field operators
$[\A](f)$ which may be smeared only with co-closed (divergence-free) 
test functions, i.e., $f$ must satisfy $\bdelta f=0$. These objects may
be interpreted as smeared gauge-equivalence classes of quantum one-form
fields: formally, 
\begin{equation}
[\A](f) \;\mbox{`}{=}\mbox{'}\; \langle \A, f\rangle\,,
\end{equation}
where $\A$ is a representative of the equivalence class $[\A]$; since
$\bdelta f=0$, we have $\langle \bd\chi,f\rangle = \langle \chi,\bdelta
f\rangle =0$ so this interpretation is indeed gauge independent. Adapting
this idea to our present setting, we start with a 
set of 
abstract objects $\{ [\A](f)\ | f\in\Omega^1_0(\M)\ \mbox{with}\ \bdelta f =
0\}$ labelled only by co-closed test one-forms. As
before, we use this set to generate a free unital $*$-algebra
over $\complex$ and define the algebra of observables $\mathfrak{A}(\M,g)$ 
to be the quotient of this algebra by the relations:
\begin{enumerate}
\item[M1.] Linearity: $[\A] \left(\alpha f_1 + \beta f_2 \right) = \alpha[\A](f_1)
+\beta[\A](f_2)$ for all $\alpha,\beta\in\complex$ and
co-closed $f_i\in\Omega^1_0(\M)$;
\item[M2.] Hermiticity: $[\A](f)^* = \A(\overline{f})$
 for all co-closed $f\in \Omega^1_0(\M)$;
\item[M3.] Field Equations: $[\A] \left( \bdelta \bd f\right)
= 0$ for all $f \in \Omega^1_0(\M)$;
\item[M4.] CCRs: $\left[[\A](f_1),[\A](f_2)\right] = -iE_0(f_1, f_2)\id$
for all co-closed $f_i\in\Omega^1_0(\M)$.
\end{enumerate}
(Note that the one-forms $f$ appearing in axiom M3
need not be co-closed.)

To see that these algebras are equivalent to those constructed
in~\cite{Dimock,Furl99}, it suffices to observe that, firstly,
the field operators of~\cite{Dimock,Furl99} certainly satisfy the
relations above (see Proposition~8 of~\cite{Dimock} and Theorem~3
of~\cite{Furl99}; note that there are notational differences) and,
secondly, that the algebras we have constructed admit no nontrivial
quotients, as may be seen by applying the theory of Sect.~7.1
in~\cite{BSZ}\footnote{For example, $\mathfrak{A}_\mass (\M,g)$ is
the infinitesimal Weyl algebra over the real symplectic space
$\Omega_0^1(\M;\real)/\ker\Delta_\mass$ (with the quotient topology)
with the addition of the $*$-operation. This has no nontrivial two-sided
$*$-ideals (see Sect.~7.1 in~\cite{BSZ}).}. 
Accordingly, our algebras are isomorphic to those 
of~\cite{Dimock,Furl99}.


\subsection{Quantisation: Hadamard states} \label{sect:QHad}

A {\em state} on a $*$-algebra $\mathfrak{A}$ is a linear functional
$\omega:\mathfrak{A}\to\complex$ which is normalised so that 
$\omega(\id) = 1$ and has the positivity property $\omega(\B^* \B) 
\geq 0$ for all $\B \in \mathfrak{A}$. However, not all states on
$\mathfrak{A}_\mass(\M,g)$ and $\mathfrak{A}(\M,g)$ are of physical 
relevance---many are insufficiently regular to permit the definition of
the stress-tensor, for example. We will focus attention on the class of
{\em Hadamard} states, for which the stress-tensor may be defined by
point-splitting techniques. The Hadamard condition was first stated rigorously 
for the scalar field by Kay and Wald \cite{Ka&Wa}; more recently,
Sahlmann and Verch \cite{Sah&Ver} have studied the Hadamard form for
wave equations with metric principal part. This does not immediately
cover either the Maxwell or Proca equations (which are not hyperbolic).
Nonetheless, we may exploit the close relationship of these equations to
the one-form Klein--Gordon equation to define the notion of Hadamard
states for these theories. 
 
To be more specific, a distribution
$W\in\left(\Omega_0^1(\M)\times\Omega_0^1(\M)\right)'$
is said to be of Hadamard form if its singular part takes a prescribed form in
a causal normal neighbourhood $\N$ of some Cauchy surface in $\M$.
This requirement is implemented by requiring that $W-H_k$ should be
$C^k$ for each $k\in\naturals$, where $H_k$ is a prescribed
sequence of distributions on $\N\times\N$. Furthermore, if $W$ is
a bisolution to $\Box+\mass^2$ (or even a bisolution modulo $C^\infty$)
then the Hadamard form `propagates' in the sense that $W$ will satisfy
the above criterion in any causal normal neighbourhood of any Cauchy surface in $\M$. 
It follows that the difference $W-W'$ between 
Hadamard form $(\Box+\mass^2)$-bisolutions $W$ and $W'$ is 
everywhere smooth on $\M\times\M$. For our purposes, the second crucial 
property of a Hadamard bisolution $W$ (first noted in the scalar case by
Radzikowski~\cite{Radz96}) is that its wave-front set is given
explicitly by
\begin{equation}
\WF(W)= \R:= \left\{ (x,k;x',-k')\in \dot{T}^*(\M\times \M)\ 
:\ (x,k)\sim(x',k') \mbox{ and } k\in\overline{V}_x^+ \right\}\,.
\label{eq:Hadamard_WFS}
\end{equation}
Here, $\dot{T}^*(\M\times \M)$ denotes the cotangent bundle of
$\M\times\M$ with its zero section excised, and
$(x,k)\sim(x',k')$ if and only if $k'$ is the parallel transport of $k$
along a null geodesic connecting $x$ and $x'$, to which $k$ is a cotangent
vector at $x$ (if $x=x'$ this degenerates to the requirement that $k=k'$
is null). We have also used 
$\overline{V}_x^+$ to denote the closed cone of future pointing
covectors at $x$.

We are now in a position to define the notion of a Hadamard state
for the Maxwell and Proca fields. Our definitions are similar to those
employed by Junker and Lled\'o~\cite{JunLle} and (particularly in the
Maxwell case) were influenced by early versions of their work~\cite{Fernando}.
\begin{description}
\item[Proca:] A state $\omega$ on $\mathfrak{A}_\mass(\M,g)$ is Hadamard 
if there exists a Hadamard form $(\Box+\mass^2)$-bisolution
$W_\mass$ such that
\begin{equation}
\omega\left( \A (f_1) \A (f_2) \right)
= W_\mass\left(f_1,\left(\id - \mass^{-2}\bd\bdelta\right)f_2\right) 
\label{eq:Proca_Hadamard}
\end{equation}
for all $f_i \in \Omega^1_0(\M )$.

\item[Maxwell:] A state $\omega$ on $\mathfrak{A}(\M,g)$
is Hadamard if there exists a Hadamard form $\Box$-bisolution
$W$ such that
\begin{equation}
\omega\left( [\A ](f_1) [\A ](f_2) \right)
= W\left(f_1,     f_2 \right) 
\label{eq:Maxwell_Hadamard}
\end{equation}
for all $f_i \in \Omega^1_0(\M )$ with $\bdelta f_i = 0$.
\end{description}

\noindent{\bf Remarks:} a) In neither case is the Hadamard form
bisolution uniquely determined by the state. \\
b) As stated, these conditions appear global in nature
and it is not clear, {\em a priori}, that any states satisfying these
definitions exist in general spacetimes. 
These concerns will be allayed in Sec.~\ref{sect:FNW}, where we show
that Hadamard states exist on general globally hyperbolic spacetimes
(subject to our usual topological restrictions on $\Sigma$). A key part of this
argument is the proof (in Appendix~\ref{appx:cons}) that it suffices
for equations~(\ref{eq:Proca_Hadamard}) and~(\ref{eq:Maxwell_Hadamard})
to hold for $f_i$ supported in a causal normal neighbourhood of a Cauchy
surface for them to hold for all $f_i$. In combination with an
explicit construction of a Hadamard state in ultrastatic spacetimes
(in Secs.~\ref{sect:proca_state} and~\ref{sect:maxwell_state}) 
this permits us to apply standard deformation
arguments~\cite{FNW81} to deduce the existence of Hadamard states in general.
\\
c) Since the two-point function $\omega^{(2)}(f,f'):=\omega(\A(f_1)\A(f_2))$ 
for a Hadamard state $\omega$ of the Proca field is
given by acting on a Hadamard $(\Box+\mass^2)$-bisolution with a partial 
differential operator, we may use Eq.~(\ref{eq:Hadamard_WFS}) and
the nonexpansion of the wave-front set under such operators to obtain
\begin{equation}
\WF( \omega^{(2)} ) \subseteq \R
\end{equation}
In the Maxwell case, the two-point function 
$\omega^{(2)}(f,f'):=\omega([\A](f_1)[\A](f_2))$
is not a distribution (because it is only defined on co-closed
test one-forms $f_i$). However, the two-point function of the field strength
$\F(f):=[\A](\bdelta f)$ is a bi-distribution on two-forms satisfying
\begin{equation}
\omega(\F(f_1)\F(f_2))=W(\bdelta f_1,\bdelta f_2) 
\end{equation}
and therefore has wave-front set contained in $\R$. \\
d) The Hadamard condition for electromagnetism was discussed
in~\cite{B&OT86,AllOtt92} in the context of a Faddeev--Popov 
method. Here, the action is modified by the addition of a gauge breaking
term and the ghost action, describing the dynamics of a complex, anticommuting scalar
field. In that context, the electromagnetic potential may be smeared
with arbitrary one-form test fields and so the corresponding two-point function is
a distributional bisolution to the massless Klein--Gordon equation (at
least in the Feynman gauge). In a Hadamard state (on the
algebra generated by the vector potential and the ghost fields), this
two-point function is required to be of Hadamard form [thus
generalising~(\ref{eq:Maxwell_Hadamard}) to non-coclosed $f_i$]; in
addition, a Ward identity is required to connect this two-point function
to that of the ghost field. This ensures that the contributions to the 
stress-energy tensor arising from the gauge-breaking and ghost actions
precisely cancel in Hadamard states.

\section{Hadamard states in ultrastatic spacetimes}
\label{sect:ultra}

To provide a concrete example of the foregoing definitions, we
now specialise to the class of ultrastatic spacetimes. This will lead to
a proof of the existence of Hadamard states for the Proca and Maxwell
fields in general globally hyperbolic spacetimes as well as providing
explicit two-point functions for particular Hadamard states which
will be used in Sect.~\ref{sect:examples}. We recall that a spacetime
$(\M,g)$ is said to be ultrastatic if $\M=\real\times\Sigma$ and $g=1\oplus -h$,
where $(\Sigma,h)$ is a smooth Riemannian manifold. As usual, we will also assume
that $\Sigma$ is compact and the homology group $H_1(\Sigma)$ is trivial. 
We will often denote a point in $\M$ by the pair $(t,\ux)$, with $t\in\real$,
$\ux\in\Sigma$. 

Our discussion in this section will proceed as follows. We begin with the
construction of a one-form Hadamard $(\Box+\mass^2)$-bisolution
$W_\mass$ on $(\M,g)$. Next, we construct Fock representations of
the Proca and Maxwell fields on $(\M,g)$ with the property that the
Fock vacuum in each case is Hadamard according to the definitions of
the preceding subsection. In each case, the $W_\mass$ plays the role of
the required Hadamard $(\Box+\mass^2)$-bisolution. We also discuss the
relationship of our procedure to the Gupta--Bleuler approach and the
method of quantisation in a fixed gauge. Finally, we explain
how these results imply the existence of Hadamard states in general
spacetimes. Some parts of the analysis are relegated to Appendix~\ref{appx:bisolution}.

\subsection{Construction of a Hadamard $(\Box+\mass^2)$-bisolution}
\label{sect:constr}

As is well-known, the Klein--Gordon equation in an ultrastatic spacetime
is readily reduced to the analysis of an elliptic eigenvalue problem.
Namely, if $\xi\in\Omega^1(\M)$ is a static one-form (i.e., independent
of the ultrastatic time parameter $t$) then $\A(t,\ux)=e^{-i\omega
t}\xi(\ux)$ solves the one-form Klein--Gordon equation if and only if
\begin{equation}
K\xi = \omega^2\xi
\label{eq:eigenprob}
\end{equation}
where the operator $K$ acts on $\xi(\ux) = \xi_0(\ux)\bd t +
\xi_\Sigma(\ux)$ ($\xi_0\in\Omega^0(\Sigma)$,
$\xi_\Sigma\in\Omega^1(\Sigma)$) by
\begin{equation}
K\xi = \left((-\Delta^s_\Sigma+\mass^2)\xi_0\right)\bd t +
(-\Delta_\Sigma+\mass^2)\xi_\Sigma\,,
\label{eq:Kdef}
\end{equation}
with $\Delta^s_\Sigma$ and $\Delta_\Sigma$ denoting the scalar and one-form
Laplace-Beltrami operators on $(\Sigma,h)$. 

To analyse $K$ further, it is helpful to have various inner product
spaces in mind. For each $p$, the space $\Omega^p(\Sigma)$ may be endowed with
a positive definite inner product
\begin{equation}
(u,v)_{\Lambda^p(\Sigma)} = \int_\Sigma \overline{u} \wedge *_\Sigma v\,,
\end{equation}
where $*_\Sigma$ is the Hodge operator on $(\Sigma,h)$. By completing
$\Omega^p(\Sigma)$ with respect to the corresponding norm we obtain
the Hilbert space $\Lambda^p(\Sigma)$ of square integrable $p$-forms 
on $\Sigma$. The direct sum $\HH=\Lambda^0(\Sigma)\oplus\Lambda^1(\Sigma)$
therefore corresponds to a Hilbert space of static one-forms, on which
the $K$ may be defined as a positive self-adjoint
operator (by using Eq.~(\ref{eq:Kdef}) to define
$K$ on $\Omega^0(\Sigma)\oplus\Omega^1(\Sigma)$ and then forming the 
Friedrichs extension). Because $\Sigma$ is compact, $K$ has purely
discrete spectrum. 

{}From a geometrical viewpoint, however, the Hilbert space inner product is not
completely natural. It is therefore convenient to introduce
an indefinite inner product on $\HH$ by
\begin{equation}
\dip{\xi}{\eta} = \int_\Sigma \overline{\xi_\mu(\ux)}\,\eta^\mu(\ux)\,
d{\rm vol}_h(\ux) = (\xi_0,\eta_0)_{\Lambda^0(\Sigma)} -
(\xi_\Sigma,\eta_\Sigma)_{\Lambda^1(\Sigma)}\,.
\end{equation}
We will denote the resulting indefinite inner product space (also known
as a {\em Krein space}) by $\KK$. In addition, we 
will say that a set $\{\xi_j: j\in J\}$ of vectors (labelled
by the elements of some set $J$) in $\KK$ is
{\em pseudo-orthonormal} if $\dip{\xi_j}{\xi_j}=\pm 1$ for each $j$ and
$\dip{\xi_j}{\xi_{j'}}=0$ for all $j\not=j'$; the set is said to be
complete if 
\begin{equation}
\eta = \sum_{j\in J} \dip{\xi_j}{\xi_j}\,\dip{\xi_j}{\eta}\,\xi_j
\label{eq:completeness}
\end{equation}
holds for all $\eta\in\KK$, with the sum converging in the topology
induced by the norm of $\HH$. We will refer to $\xi\in\KK$ as
`timelike', `spacelike' or `null' depending on whether $\dip{\xi}{\xi}$
is positive, negative or zero, respectively. 

The following result is proved in Appendix~\ref{appx:bisolution}.
\begin{Theorem}
\label{thm:Hadbisoln}
Let $\xi_j$ ($j\in J$) be a complete, pseudo-orthonormal basis for $\KK$
such that $K\xi_j = \omega_j^2\xi_j$ ($\omega_j\ge 0$) and
set $\A_j(t,\ux)=e^{-i\omega_j t}\xi_j(\ux)$. Then
\begin{equation}
W_\mass(f_1,f_2)=-\sum_{j\in J:\omega_j>0} \frac{1}{2\omega_j}
\dip{\xi_j}{\xi_j}\langle \A_j,f_1\rangle\,\langle
\overline{\A_j},f_2\rangle\,, \qquad(f_i\in\Omega^1(\Sigma))
\end{equation}
defines a one-form Hadamard $(\Box+\mass^2)$-bisolution $W_\mass$ on
$(\M,g)$ which is independent of the particular basis chosen. For
$\mass>0$, we have
\begin{equation}
W_\mass(f_1,f_2) - W_\mass(f_2,f_1) = -iE_\mass(f_1,f_2)
\label{eq:antisymmpart}
\end{equation}
for all $f_i\in\Omega_0^1(\M)$; in the case $\mass=0$, this 
holds provided the $f_i$ are both co-closed. 
\end{Theorem}

Our task is now to analyse the eigenproblem~(\ref{eq:eigenprob}).
We may identify various families of eigenfunctions which are 
more or less convenient for different purposes and which may be combined
to form complete pseudo-orthonormal bases (again, different bases
are useful in different contexts). A typical eigenfunction will be
denoted $\xi(\lambda,j)$, where $\lambda$ labels the family to which it
belongs and $j$ (which labels eigenfunctions within families) 
takes values in a labelling set $J(\lambda)$. Because $K$ is
manifestly positive, the eigenvalues may be expressed as the
squares of nonnegative quantities $\omega(\lambda,j)$: thus
$K\xi(\lambda,j)=\omega(\lambda,j)^2\xi(\lambda,j)$. The
corresponding Klein--Gordon solution is
\begin{equation}
\A(\lambda,j)(t,\ux)=e^{-i\omega(\lambda,j)t}\xi(\lambda,j)(\ux)\,.
\end{equation}
We begin by identifying three eigenfunction families which together
form a simultaneously $\KK$-pseudo-orthonormal and $\HH$-orthonormal
basis. The analysis of Eq.~(\ref{eq:eigenprob}) is considerably simplified by
the fact that it decouples
into the two equations
\begin{eqnarray}
(-\Delta_\Sigma^s+\mass^2)\xi_0 &=& \omega^2\xi_0 \label{eq:epa} \\
(-\Delta_\Sigma+\mass^2)\xi_\Sigma &=& \omega^2\xi_\Sigma\,.
\end{eqnarray}
Accordingly, choosing $\varphi_j$ ($j\in J(S)$) to label a complete
orthonormal basis of $(-\Delta_\Sigma^s+\mass^2)$-eigenfunctions for 
$\Lambda^0(\Sigma)$ with corresponding eigenvalues
$\omega(S,j)^2$, we may identify a family of ``scalar'' $K$-eigenfunctions given by
\begin{equation}
\xi(S,j) = \varphi_j\,\bd t \qquad (j\in J(S))\,.
\end{equation}
(By elliptic regularity, each $\varphi_j$ is in fact smooth.) Clearly
the $\xi(S,j)$ are pseudo-orthonormal and `timelike' in $\KK$, as
well as being $\HH$-orthonormal.  
For future reference, we note that there is a unique spatially constant
eigenfunction ${\rm vol}_h(\Sigma)^{-1/2}\bd t$ with eigenvalue $\mass^2$. 
Writing $\Phi_j(t,\ux) = e^{-i\omega(S,j) t}\varphi_j(\ux)$, the
corresponding one-form positive frequency modes are
\begin{equation}
\A(S,j) = \Phi_j\, \bd t\,.
\label{eq:scalar_def}
\end{equation}

The remaining pseudo-orthogonal modes must have vanishing $\bd t$
component; that is, they must lie in the `spacelike' subspace
$\Lambda^1(\Sigma)$ of $\HH$. On this
subspace, the inner products of $\HH$ and $\KK$ differ by an overall
sign only, so $\KK$-pseudo-orthonormality and $\HH$-orthonormality are
again identical. Since $H_1(\Sigma)$ is trivial, the Hodge
decomposition (see, e.g., Prop.~11.7 in~\cite{CFKS}) gives $\Lambda^1(\Sigma)=
\overline{\bd_\Sigma\Omega^0(\Sigma)}\oplus \overline{\bdelta_\Sigma\Omega^2(\Sigma)}$, where the
decomposition is orthogonal with respect to 
$(\cdot,\cdot)_{\Lambda^1}$ and hence $\dip{\cdot}{\cdot}$, and the bars
denote closure. As is easily verified, the Laplacian $\Delta_\Sigma$ is
block diagonal with respect to this decomposition, thus enabling us to
seek eigenfunctions within each subspace in turn. If the restriction
on $H_1(\Sigma)$ were removed, there would be a third subspace, 
consisting of harmonic forms (i.e., the kernel of $\Delta_\Sigma$). 
In the Proca case, the resulting modes lead to additional terms in various expansions
given below and in the quantum inequality, but do not significantly
alter the formalism. By contrast, the incorporation of
the harmonic modes in the Maxwell case raises nontrivial issues, to which we hope to return
elsewhere.

Now, the ``longitudinal'' subspace $\overline{\bd_\Sigma\Omega^0(\Sigma)}$
is clearly spanned by
the set of non-vanishing vectors of the form $\bd\xi(S,j)$ ($j\in
J(S)$); owing to the relation $\Delta_\Sigma\bd_\Sigma=\bd_\Sigma\Delta^s_\Sigma$,
these must also be eigenvectors for $-\Delta_\Sigma+\mass^2$ with
eigenvalue $\omega(L,j)^2=\omega(S,j)^2$. 
As the only vanishing vector of this form is obtained from the
spatially constant mode, the appropriate labelling set is 
$J(L)=\{j\in J(S): \omega(S,j)>\mass\}$. 
Furthermore, the calculation
\begin{equation}
\dip{\bd_\Sigma\varphi_{j}}{\bd_\Sigma\varphi_{j'}}  = 
-(\bd_\Sigma\varphi_{j},\bd_\Sigma\varphi_{j'})_{\Lambda^1(\Sigma)} =
-(\varphi_{j},\bdelta_\Sigma\bd_\Sigma\varphi_{j'})_{\Lambda^1(\Sigma)}
= -(\omega(S,j)^2-\mass^2)\delta_{jj'}
\end{equation}
shows that the appropriately normalised longitudinal eigenfunctions are
\begin{equation}
\xi(L,j)=
(\omega(S,j)^2-\mass^2)^{-1/2}\bd_\Sigma\varphi_j\,,\qquad(j\in J(L))\,.
\end{equation}
The corresponding Klein--Gordon modes may also be expressed in the form
\begin{equation}
\A(L,j) = \frac{\bd \Phi_j +
i\omega(L,j)\Phi_j\, \bd t}{\sqrt{\omega(L,j)^2-\mass^2}}\,.
\label{eq:axial_def}
\end{equation}
All remaining pseudo-orthonormal modes must lie in the coexact subspace
$\bdelta_\Sigma\Omega^2(\Sigma)$. We will refer to these as the transverse
modes $\xi(T,j)$ with labelling set $J(T)$ and eigenfrequencies
$\omega(T,j)$; they are necessarily `spacelike'. 
In general, the $\omega(T,j)$'s will be distinct
from the $\omega(S,j)$'s (even in situations of high symmetry such as
the Einstein spacetime---see Sect.~\ref{sect:Einstein}).

Applying Theorem~\ref{thm:Hadbisoln}, we obtain a Hadamard bisolution
\begin{equation}
W_M(f_1,f_2) = -\sum_{\lambda\in\{S,L,T\}}
\sum_{j\in J(\lambda)}\frac{s(\lambda)}{2\omega(\lambda,j)}
\langle \A(\lambda,j),f_1\rangle\,\langle
\overline{\A(\lambda,j)},f_2\rangle\,,
\end{equation}
where $s(S)=1$, $s(L)=s(T)=-1$. 

\subsection{The Proca field}
\label{sect:proca_state}

We now construct a state on $\mathfrak{A}_\mass(\M,g)$
whose two-point function is related to $W_\mass$ by
Eq.~(\ref{eq:Proca_Hadamard}) and is therefore Hadamard. We 
begin by finding an expression for
$W_\mass(f_1,(\id-M^{-2}\bd\bdelta)f_2)$. 
To this end, we first note that the transverse modes $\xi(T,j)$, being
coexact, are necessarily coclosed: $\bdelta_\Sigma\xi(T,j)=0$. 
It follows that $\bdelta\A(T,j)=0$, so transverse modes therefore
solve the Proca equation. The same cannot be said of the scalar and
longitudinal modes, for which 
\begin{equation}
\bdelta\A(S,j) =i\omega(S,j) \Phi_j
\end{equation}
for all $j\in J(S)$, and
\begin{equation}
\bdelta\A(L,j) =-\sqrt{\omega(S,j)^2-\mass^2} \Phi_j
\end{equation}
for all $j\in J(L)$. It is convenient to change basis, introducing
\begin{eqnarray}
\xi(P,j)&=&\frac{1}{\mass}
\left[\omega(S,j)\xi(L,j)-i\sqrt{\omega(S,j)^2-\mass^2}
\xi(S,j)
\right]\\
\xi(G,j)&=&\frac{1}{\mass} \left[\sqrt{\omega(S,j)^2-\mass^2}
\xi(L,j)-i\omega(S,j)\xi(S,j) 
\right]
\end{eqnarray}
with labelling sets $J(P)=J(G)=J(L)$. The two sets are
pseudo-orthonormal, with $\xi(P,j)$ and $\xi(G,j)$ 
being `spacelike' and `timelike' respectively. By construction, 
the ``scalar Proca'' modes
$\A(P,j)(t,\ux)=e^{-i\omega(P,j)t}\xi(P,j)(\ux)$ are coclosed
and therefore obey the Proca equation. We also note the expressions
\begin{eqnarray}
\A (P ,j)&=&\frac{1}{\mass} \left[ \omega(S,j) \A(L,j)-i\sqrt{\omega(S,j)^2-\mass^2}
\A(S,j)\right]\,,\nonumber\\
&=&
\frac{i\mass}{\sqrt{\omega(S,j)^2-\mass^2}}
\left[ \Phi_j\, \bd t-i\frac{\omega(S,j)}{\mass^2}\,\bd \Phi_j\right]\,.
\label{eq:scalar_proca_def}
\end{eqnarray}
On the other hand, 
the corresponding ``gradient'' modes $\A(G,j)$ have
\begin{equation}
\A(G,j) = \mass^{-1}\bd\Phi_j 
\end{equation}
and therefore have nonvanishing coderivative
\begin{equation}
\bdelta \A(G,j) = \mass\Phi_j\,.
\end{equation}

{}From Theorem~\ref{thm:Hadbisoln}, we know that $W_\mass$ is basis
independent and may therefore be expressed in terms of the
$\A(P,j)$, $\A(G,j)$, $\A(T,j)$ and the unique spatially constant mode. 
Of these, the scalar Proca and transverse modes are left invariant by
the operator $(\id-\mass^{-2}\bd\bdelta)$, while the spatially constant
mode and the gradient modes are annihilated. Accordingly, with
$W_\mass$ expressed in the new basis, we find
\begin{equation}
W_\mass((\id-M^{-2}\bd\bdelta)f_1,f_2) = 
W_\mass(f_1,(\id-M^{-2}\bd\bdelta)f_2)
=\sum_{\lambda\in\{T,P\}}
\sum_{j\in J(\lambda)}\frac{1}{2\omega(\lambda,j)}
\langle \A(\lambda,j),f_1\rangle\,\langle
\overline{\A(\lambda,j)},f_2\rangle\,.
\end{equation}
As a consequence of the first equality and Eq.~(\ref{eq:antisymmpart})
we see that
\begin{equation}
W_\mass(f_1,(\id-M^{-2}\bd\bdelta)f_2)-
W_\mass(f_2,(\id-M^{-2}\bd\bdelta)f_1)=
-iE_\mass(f_1,(\id-M^{-2}\bd\bdelta)f_2)=-i\Delta_\mass(f_1,f_2)
\end{equation}
which in turn leads to the expansion
\begin{equation}
\Delta_\mass(f_1,f_2) = \sum_{\lambda\in\{T,P\}}
\sum_{j\in J(\lambda)}\frac{1}{2\omega(\lambda,j)}
\left(\langle \A(\lambda,j),f_1\rangle\,\langle
\overline{\A(\lambda,j)},f_2\rangle - 
\langle \A(\lambda,j),f_2\rangle\,\langle
\overline{\A(\lambda,j)},f_1\rangle\right)\,.
\end{equation}

To complete the analysis, we define smeared fields
\begin{equation}
\A(f)= \sum_{\lambda\in\{T,P\}}
\sum_{j\in J(\lambda)}\frac{1}{\sqrt{2\omega(\lambda,j)}}
\left(\langle \A(\lambda,j),f\rangle\,a(\lambda,j)+ 
\langle \overline{\A(\lambda,j)},f\rangle\,a(\lambda,j)^*\right)
\end{equation}
on the Fock space $\FF_\mass$ generated by 
operators $a(\lambda,j)$ obeying
$[a(\lambda,j),a(\lambda',j')^*]=\delta_{\lambda\lambda'}\delta_{jj'}\id$. 
It is now easily verified that the $\A(f)$ provide a representation of 
axioms P1--P4 on (a dense domain in) $\FF_\mass$. Thus the Fock vacuum
$\vert 0\rangle$ defines a state on $\mathfrak{A}_\mass(\M,g)$ with two-point
function
\begin{equation}
\langle 0 \vert \A(f_1)\A(f_2)\vert 0\rangle=
W_\mass(f_1,(\id-M^{-2}\bd\bdelta)f_2)\,.
\end{equation}
Accordingly, this state is Hadamard. Acting on the vacuum with smeared
fields, we obtain a dense set of Hadamard states in $\FF_\mass$. 

\subsection{The Maxwell field}
\label{sect:maxwell_state}

We now repeat the above analysis for the Maxwell field. In this
instance, we begin by finding an expression for $W_0(f_1,f_2)$ for
co-closed $f_i$. Since the $\A(P,j)$ and $\A(G,j)$ diverge as $\mass\to
0$, we revert to the scalar, longitudinal and transverse modes of
Sect.~\ref{sect:constr}. Now, from Eqs.~(\ref{eq:scalar_def}) and~(\ref{eq:axial_def}), 
it is simple to show $\langle \A(L,j) ,F \rangle = i \langle \A(S,j),F \rangle$
for all co-closed $F$.  This has the effect that the contributions to
$W_0(f_1,f_2)$ from the scalar and longitudinal modes cancel and so
only has contributions from the transverse modes:
\begin{equation}
W_0(f_1,f_2) = 
\sum_{j\in J(T)}\frac{1}{2\omega(T,j)}
\langle \A(T,j),f_1\rangle\,\langle
\overline{\A(T,j)},f_2\rangle\,,
\end{equation}
for any co-closed $f_i$. Defining 
\begin{equation}
[\A](f)= 
\sum_{j\in J(T)}\frac{1}{\sqrt{2\omega(T,j)}}
\left(\langle \A(T,j),f\rangle\,a(T,j)+ 
\langle \overline{\A(T,j)},f\rangle\,a(T,j)^*\right)
\label{eq:em_fieldops}
\end{equation}
on the Fock space $\FF_0$ generated by 
operators $a(T,j)$ obeying
$[a(T,j),a(T,j')^*]=\delta_{jj'}\id$, we may again verify that 
the $[\A](f)$ provide a representation of 
axioms M1--M4 on (a dense domain in) $\FF_0$. Thus the Fock vacuum
$\vert 0\rangle$ defines a state on $\mathfrak{A}(\M,g)$ with two-point
function
\begin{equation}
\langle 0 \vert [\A](f_1)[\A](f_2)\vert 0\rangle=
W_0(f_1,f_2)
\end{equation}
for all co-closed $f_i$ and is therefore Hadamard, as
are states obtained from it by acting with polynomials in smeared fields.

\subsection{Comparison with other quantisation methods}

It is instructive to compare our approach to two more familiar
methods of quantisation, in particular, the
Gupta--Bleuler formalism~\cite{Gupta50,Gupta} and (for electromagnetism) 
quantisation in a fixed gauge. A detailed presentation of these
methods for electromagnetic fields in static spacetimes can be found
in~\cite{Pfen02}. We will confine our remarks to the Maxwell case, but
analogous comments apply to the Proca field.  

As is well-known, the Gupta--Bleuler procedure starts
by adding a gauge breaking term to the Maxwell Lagrangian. By choosing
the coefficient of this term appropriately, the field equations become precisely those of
the massless Klein--Gordon equation. (This is sometimes known as the
`Feynman gauge'.) This equation is then quantised
using an indefinite inner product space (effectively the symmetric
Fock space $\FF(\KK)$ over our Krein space $\KK$). The physical Hilbert space is
then selected by the requirement that they be annihilated by the positive frequency
part of the divergence of the resulting field operators. 

In our approach the various ingredients of the Gupta--Bleuler formalism
appear in a different way. We start from the Hadamard bisolution $W_0$,
which is in fact the two-point function of the static vacuum of the
Gupta--Bleuler theory in the indefinite space $\FF(\KK)$. In
our approach, however, the 
restriction to the physical state space occurs at the `one-particle
level': the Hadamard condition and the restriction to co-closed 
test one-forms pick out a preferred Hilbert subspace of
$\KK$, namely, the subspace $\HH_{\rm phys}$ 
spanned by the transverse modes $\xi(T,j)$. 
The quantum fields are then given immediately as operators on the symmetric Fock
space $\FF_0=\FF(\HH_{\rm phys})$ over this Hilbert space. 
The key advantage of our approach over the Gupta--Bleuler method is, of
course, that it is not tied
to a particular notion of `positive frequency' and is applicable in
general globally hyperbolic spacetimes.

An alternative to the Gupta--Bleuler approach is to fix a gauge
from the outset. Employing the Coulomb gauge~\cite{Pfen02}, we would seek
modes $\A_j$ obeying $\bdelta\A_j=0$ with vanishing $\bd t$ component.
These modes are easily seen to constitute the transverse family
$\A(T,j)$ and we essentially recover the field operators
of Eq.~(\ref{eq:em_fieldops}), but without the restriction to co-closed
$f_i$. By contrast, no gauge is ever chosen in our approach and the
transverse modes emerge naturally from the Hadamard condition in
combination with the restriction to co-closed $f_i$.

\subsection{Existence of Hadamard states in globally hyperbolic spacetimes}
\label{sect:FNW}

Once the existence of Hadamard states has been established in
ultrastatic spacetime, arguments due to Fulling, Narcowich and Wald~\cite{FNW81}
may be used to deduce their existence in general globally
hyperbolic spacetimes obeying our usual topological restrictions. We illustrate
this for the Proca field, indicating the slight differences required in
the Maxwell case. 

Let $(\M,g)$ and $(\M',g')$ be globally hyperbolic spacetimes
and suppose that there is an isometry $\psi:\N\to\N'$ between causal
normal neighbourhoods $\N$ and $\N'$ of Cauchy surfaces in $\M$ and $\M'$
respectively. Then any state $\omega$ on
$\mathfrak{A}_\mass(\M,g)$ determines a state $\omega'$ on 
$\mathfrak{A}_\mass(\M',g')$ as follows: given any
$f_1,\ldots,f_n\in\Omega_0^1(\M')$ choose $\widetilde{f}_1,\ldots,
\widetilde{f}_n\in\Omega_0^1(\N')$ such that $\widetilde{f}_k-f_k\in 
(-\bdelta\bd+\mass^2)\Omega_0^1(\M')$. (The existence of such
$\widetilde{f}_k$ follows from Proposition~\ref{lem:eom_test}(a).) Now
define 
\begin{equation}
\omega'(\A'(f_1)\cdots\A'(f_n)) = 
\omega(\A(\psi^*\widetilde{f}_1)\cdots\A(\psi^*\widetilde{f}_n))\,,
\end{equation}
where $\psi^*$ denotes the pull-back, $\psi^*f=f\circ\psi$ and
we have used $\A'(f)$ to denote field operators in
$\mathfrak{A}_\mass(\M',g')$. 
It is easily verified that $\omega'$ is a state on
$\mathfrak{A}_\mass(\M',g')$. Moreover, if $\omega$ is Hadamard then the
isometry ensures that 
$\omega'(\A(f_1)\A(f_2)) =W(f_1,(\id-\mass^{-1}\bd\bdelta)f_2)$ for all $f_i\in\Omega_0^1(\N')$,
where $W$ is a $(\Box+\mass^2)$-bisolution Hadamard bisolution on
$\N\times\N$. It then follows from Theorem~\ref{thm:con} that $\omega'$ is
Hadamard. 

Given a globally hyperbolic spacetime $(\M,g)$, we may now employ
a construction described in~\cite{FNW81} to
obtain a globally hyperbolic spacetimes $(\M',g')$ and $(M'',g'')$ so
that (i)~$(\M'',g'')$ is ultrastatic; (ii)~$\M'$ contains Cauchy surfaces
$\Sigma'_1$ and $\Sigma'_2$ with causal normal neighbourhoods $\N'_1$
and $\N'_2$ so that $\N'_1$ (respectively, $\N'_2$) is isometric to
a causal normal neighbourhood of a Cauchy surface in $(\M,g)$
(respectively, $(\M'',g'')$). As all the Cauchy surfaces involved are
therefore homeomorphic, $(\M',g')$ and $(\M'',g'')$ will obey our
topological restrictions if $(\M,g)$ does. Starting with a Hadamard
state on the ultrastatic spacetime $(\M'',g'')$ we may induce Hadamard
states on $(\M',g')$ and hence on $(\M,g)$. Thus
$\mathfrak{A}_\mass(\M,g)$ admits Hadamard states. 

An analogous argument applies in the Maxwell case. The only 
differences are that all test functions must now be co-closed, 
we use part~(b) of Proposition~\ref{lem:eom_test} instead of part~(a),
and use the appropriate form of the Hadamard condition.

\section{A Quantum Weak Energy Inequality}
\label{sect:QWEI}

We now proceed to prove our quantum inequality. Let $\tau\mapsto\gamma(\tau)$
be a smooth timelike curve in $\M$ parametrised by its proper time $\tau\in\real$, 
and let
$\Gamma$ be a tubular neighbourhood of $\gamma$. Inside $\Gamma$
let $\{ v_i^\rho | i = 0,1,2,3\}$ be an orthonormal frame obeying 
$g^{\rho\sigma}= \eta^{ij} v_i^\rho v_j^\sigma$ and with the property
that the restriction of $v_0$ to $\gamma$ is the four-velocity
$u^\sigma(\tau)=(d\gamma(\tau)/d\tau)^\sigma$ of the
curve: $v_0^\sigma(\gamma(\tau)) =u^\sigma(\tau)$. (See~\cite{Fe&V02}
for an explicit construction of such a frame.)

The first step is to construct the quantised energy density measured
along the curve. The starting point is the 
classical stress-energy tensor, which for the Proca and Maxwell fields,
takes the form
\begin{equation}
T_{\mu\nu}= \frac{1}{4} g_{\mu\nu} \F_{\rho\sigma} \F^{\rho\sigma}
-\F_{\mu\rho} {\F_\nu}^\rho 
+\mass^2\left(\A_\mu \A_\nu - \frac{1}{2}g_{\mu\nu} \A_\rho \A^\rho
\right),
\end{equation}
with $\mass=0$ in the Maxwell case. The energy density along the curve
is given by $T_{\mu\nu}(\gamma(\tau))u^\mu(\tau)u^\nu(\tau)$; we may 
extend this quantity off the curve by defining 
$\T(x) =T_{\mu\nu}(x)v_0^\mu(x) v_0^\nu(x)$. A little manipulation
shows that 
\begin{equation}
\T(x)=\frac{1}{4} \sum_{i=0}^3 \sum_{j=0}^3 \left(v_i^\rho v_j^\sigma
\F_{\rho\sigma}\right)^2 + \frac{1}{2}\mass^2 \sum_{i=0}^3 \left(v_i^\rho
\A_\rho\right)^2.
\end{equation}
We may rewrite $\T$ in terms of forms with the aid of the one-form basis
$\{V^i:i=0,1,2,3\}$ dual to the $v_i$, given by $V_\mu^i =
\eta^{ij}g_{\mu\nu}v_j^\nu$. Writing $W^{ij}=V^i\wedge V^j$, we find
\begin{equation}
\T(x) =\frac{1}{4}\sum_{i=0}^3\sum_{j=0}^3 \left[*\left(\F\wedge* W^{ij}\right)\right]^2+
\frac{1}{2}\mass^2\sum_{i=0}^3 \left[*\left(\A\wedge * V^i\right)\right]^2\,,
\end{equation}
which is clearly the restriction to the diagonal $x'=x$ of
\begin{widetext}
\begin{equation}
T(x,x') = \frac{1}{4}\sum_{i=0}^3 \sum_{j=0}^3
\left.\left[*\left(\F\wedge* W^{ij}\right)\right]\right|_{x} 
\left.\left[*\left(\F\wedge* W^{ij}\right)\right]\right|_{x'} 
+
\frac{1}{2}\mass^2\sum_{i=0}^3 
\left.\left[*\left(\A\wedge * V^i\right)\right]\right|_{x}
\left.\left[*\left(\A\wedge * V^i\right)\right]\right|_{x'}
\,.
\end{equation}
The function $T(x,x')$
provides a point-split classical energy density, related to the
true energy density on the curve $\gamma$ by
$\T(\gamma(\tau))=T(\gamma(\tau),\gamma(\tau))$. Moreover, for
$f,f'\in\D(\M)$ we see that
\begin{eqnarray}
\int_{\M\times\M} T(x,x') f(x) f'(x') d{\rm vol}_g(x)\,d{\rm vol}_g(x')
&=& \frac{1}{4}\sum_{i=0}^3 \sum_{j=0}^3 \langle \F,W^{ij}f\rangle \langle \F,W^{ij}f'\rangle
+\frac{1}{2}M^2 \sum_{i=0}^3 \langle \A,V^i f\rangle \langle \A,V^i f'\rangle
\nonumber\\
&=& \frac{1}{4}\sum_{i=0}^3 \sum_{j=0}^3 \langle \A,\bdelta(W^{ij}f)\rangle \langle \A,\bdelta(W^{ij}f')\rangle
+\frac{1}{2}M^2 \sum_{i=0}^3 \langle \A,V^i f\rangle \langle \A,V^i
f'\rangle\,.
\nonumber\\
&&
\end{eqnarray}
\end{widetext}
The advantage of this reformulation is that it is easily quantised: given a
Hadamard state $\omega$ on $\mathfrak{A}_\mass (\M,g)$ for the Proca field,
or $\mathfrak{A}(\M,g)$
for the Maxwell field we simplify replace occurrences of $\langle
\A,F\rangle\langle \A,F'\rangle$ with the two-point function $\omega^{(2)}(F,F')$ thus
obtaining a scalar bi-distribution $\langle T
\rangle_\omega\in \D'(\M\times\M)$ given by
\begin{equation}
\langle T \rangle_\omega (f,f') =\frac{1}{4} 
\sum_{i=0}^3 \sum_{j=0}^3 \omega^{(2)}\left(\bdelta (W^{ij}f),\bdelta (W^{ij}f')\right)
+
\frac{1}{2}\mass^2\sum_{i=0}^3 \omega^{(2)}\left(V^i f,V^i f'\right)
\,,
\label{eq:smeared_unrenorm_T}
\end{equation}
(Note that in the Maxwell case, the above expression reduces to its
first term, in which the arguments of the two-point function are co-closed.) 

Now suppose that a reference Hadamard state $\omega_0$ is specified.
Since the difference between Hadamard form $(\Box+\mass^2)$-bisolutions
is smooth,  
\begin{equation}
\langle :T:\rangle_\omega =\langle  T \rangle_\omega - \langle  T
\rangle_{\omega_0}
\label{eq:splitnoedens}
\end{equation} 
is easily seen to be a smooth function on
$\M\times\M$. Thus we may `unsplit points' to define the normal ordered energy density along $\gamma$
by
\begin{equation}
\langle:\rho:\rangle_\omega(\tau) = \langle :T:\rangle_\omega(\gamma(\tau),\gamma(\tau))\,.
\label{eq:noedens}
\end{equation}

We will need three further observations. First, $\langle
:T:\rangle_\omega$ is symmetric as a consequence of the commutator axiom
P4/M4. Second, $\langle T \rangle_\omega$ is
a distribution of {\em positive type}, i.e., $\langle T
\rangle_\omega(\overline{f},f)\ge 0$ for all $f\in\D(\M)$. This is
ultimately a consequence of the positivity property $\omega(\B^*\B)\ge
0$ and the hermiticity axiom P2/M2. Third, the wave-front set of 
$\langle T \rangle_\omega$ obeys
\begin{equation}
\WF(\langle T \rangle_\omega) \subseteq \R\,.
\end{equation}
In the Proca case, this is a straightforward consequence of the 
non-expansion of the wave-front set under partial differential
operators. In the electromagnetic case, the key point is that 
$(F,G)\mapsto \omega^{(2)}(\bdelta F,\bdelta G)$ is a distribution (the
two-point function of the field strength) whose wave-front set is known
to be contained in $\R$.

Now consider the the pull-back $\varphi^*\langle T \rangle_\omega$
of the bi-distribution $\langle T \rangle_\omega$ 
induced by the smooth map $\varphi(\tau,\tau')=\left(\gamma(\tau),
\gamma(\tau')\right)$. This quantity is formally the unrenormalised energy density, with
points split along $\gamma$. We now claim that $\varphi^*\langle T 
\rangle_\omega$ is a well defined distribution of positive type.
To verify this, we first note that $^t\varphi'(\tau,\tau'): T^*_{(\gamma(\tau),
\gamma(\tau'))}(\M\times\M)\rightarrow\real$ is the linear map
\begin{equation}
^t\varphi'(\tau,\tau'):(k,k')\mapsto (u^\rho(\tau)k_\rho,
u^{\sigma'}(\tau')k'_{\sigma'})
\end{equation}
from which it follows that $\varphi$ has the following set of normals 
\begin{equation}
N_\varphi=\left\{ (\gamma(\tau),k;\gamma(\tau'),k')\in T^*(\M\times\M) :
u^\rho(\tau)k_\rho=u^{\sigma'}(\tau')k'_{\sigma'}=0\right\}\,.
\end{equation}
Second, we note that all the covectors appearing in $\WF(\langle T
\rangle_\omega)$ are null, and therefore cannot annihilate any nonzero
timelike vector. Accordingly, the intersection $\WF(\langle T
\rangle_\omega)\cap N_\varphi$ is empty and the pull-back $\varphi^*\langle T
\rangle_\omega$ exists by Theorem~2.5.11$'$ in~\cite{Hormander_FIO1} 
(in which the set of normals is also defined). Theorem~2.2 of~\cite{Fews00} guarantees that
it inherits the positive type property.

Theorem~2.5.11$'$ in~\cite{Hormander_FIO1} also asserts that the
wave-front set of $\varphi^* \langle T \rangle_\omega$ obeys
$\WF(\varphi^* \langle T \rangle_\omega) \subset \varphi^*\WF(\langle
T \rangle_\omega)$.
Thus, we have $(\tau,\zeta;\tau',-\zeta') \in \varphi^*\WF(\langle
T \rangle_\omega)$ only if
\begin{equation}
(\zeta,-\zeta') = \left(^t\varphi'(\tau,\tau')\right)(k,-k')
=(u^\rho(\tau)k_\rho,-u^{\sigma'}(\tau')k'_{\sigma'})\,.
\end{equation}
Since the vectors $u^\sigma(\tau)$, $u^{\sigma'}(\tau')$
and the covectors $k_\sigma$, $k'_{\sigma'}$ are all future pointing
their contractions will always be positive. It will therefore
be the case that $\zeta,\zeta'>0$.

Summarising, $\varphi^*\langle T \rangle_\omega$ is a well defined
distribution in $\D'(\real^2)$, which is of positive type and has a wave-front
set obeying
\begin{equation}
\WF(\varphi^* \langle T \rangle_\omega)\subset\left\{ (\tau,\zeta;\tau',-\zeta')
|\, \zeta,\zeta'>0 \right\}\,.
\end{equation}
We are now in a position to state and prove our
quantum weak energy inequality. The proof is in fact essentially
identical to that given for the scalar field in~\cite{Fews00}; it is
given here for completeness. 

\begin{widetext}
\begin{Theorem}
Let $\omega$ and $\omega_0$ be Hadamard states on $\mathfrak{A}_\mass (\M,g)$ for the Proca field,
or $\mathfrak{A}(\M,g)$ for the electromagnetic field. 
Define the point-split normal ordered energy density $\langle :T:\rangle_\omega$ relative
to $\omega_0$ by Eq.~(\ref{eq:splitnoedens}) and the normal ordered
energy density $\langle:\rho:\rangle_\omega(\tau)$ by Eq.~(\ref{eq:noedens}). Then 
the quantum inequality
\begin{equation}
\int d\tau (g(\tau))^2 \langle:\rho:\rangle_\omega(\tau)
\geq -\int_0^\infty
\frac{d\alpha}{\pi} \left[ (g\otimes g) \varphi^*\langle T \rangle_{\omega_0}\right]^\wedge
(-\alpha,\alpha)
\label{eq:the_QI}
\end{equation}
holds and the right-hand side is finite for all real-valued
$g\in C_0^\infty(\real)$.
\end{Theorem}
\end{widetext}
\proof
Letting $g\in C_0^\infty(\real)$
be a real valued function of proper time $\tau$, we have
\begin{eqnarray}
\int d\tau (g(\tau))^2 \langle:\rho:\rangle_\omega(\tau)
&=&\int d\tau\ (g(\tau))^2\varphi^*\langle :T:\rangle_\omega (\tau,\tau)\nonumber\\ 
&=&\int d\tau \int d\tau'\ g(\tau) g(\tau') \delta(\tau-\tau')
\varphi^*\langle :T:\rangle_\omega (\tau,\tau')\nonumber\\
&=&\int_{-\infty}^\infty \frac{d\alpha}{2\pi} 
\int d\tau \int d\tau'\ g_{-\alpha}(\tau) g_\alpha(\tau')
\varphi^*\langle :T:\rangle_\omega (\tau,\tau')\nonumber\\
&=&\int_{-\infty}^\infty \frac{d\alpha}{2\pi}\ 
\varphi^*\langle :T:\rangle_\omega (g_{-\alpha}\otimes g_\alpha)\nonumber\\
&=&\int_{0}^\infty \frac{d\alpha}{\pi}\ 
\varphi^*\langle :T:\rangle_\omega (\overline{g_\alpha}\otimes g_\alpha)
\end{eqnarray}
where $g_\alpha(\tau)= g(\tau)e^{i\alpha\tau}$ and we used the symmetry
property $\varphi^*\langle :T:\rangle_\omega (\overline{g_\alpha}\otimes g_\alpha)=
\varphi^*\langle :T:\rangle_\omega (g_\alpha\otimes\overline{g_\alpha})$ in the last
step. Using the definition of $\langle :T:\rangle_\omega$ and the fact
that $\varphi^*\langle T\rangle_\omega$ is of positive type we have
\begin{equation}
\int d\tau\ (g(\tau))^2\varphi^*\langle :T:\rangle_\omega (\tau,\tau) \geq
-\int_{0}^\infty \frac{d\alpha}{\pi}\ 
\varphi^*\langle T \rangle_{\omega_0} (\overline{g_\alpha}\otimes g_\alpha).
\end{equation}

Setting $e_{(\alpha,\alpha')}(\tau,\tau') = e^{i(\alpha\tau+\alpha'\tau')}$, 
we note that the integrand in the above expression may be rewritten as
\begin{eqnarray}
\varphi^*\langle T\rangle_{\omega_0} (\overline{g_\alpha}\otimes g_\alpha)
&=&\varphi^*\langle T\rangle_{\omega_0} \left( (g \otimes g) e_{(-\alpha,
\alpha)}\right)\nonumber\\
&=& \left[(g\otimes g)\varphi^*\langle T\rangle_{\omega_0}\right](e_{(-\alpha,
\alpha)})\nonumber\\ 
&=& \left[(g\otimes g)\varphi^*\langle T\rangle_{\omega_0}\right]^\wedge(-\alpha,
\alpha)
\end{eqnarray} 
by our definition for the Fourier transform. Thus we obtain Eq.~(\ref{eq:the_QI}).

The fact that the right hand side of Eq.~(\ref{eq:the_QI}) is convergent
follows from an analysis of the wave-front set. From the definition
of the wave-front set, it is obvious that
\begin{equation}
\WF\left((f_1\otimes f_2)\varphi^*\langle T\rangle_{\omega_0}\right)
\subseteq \WF\left(\varphi^*\langle T\rangle_{\omega_0}\right)
\end{equation}
for any $f_i \in C_0^\infty(\real)$. Thus the singular
directions for $(f_1\otimes f_2)\varphi^*\langle T\rangle_{\omega_0}$
are contained in $\{(\zeta,-\zeta')|\,\zeta,\zeta'>0\}$. In consequence,
the Fourier transform 
$\left[(g\otimes g)\varphi^*\langle T\rangle_{\omega_0}\right]^\wedge(-\alpha,
\alpha)$ decays rapidly as $\alpha\to+\infty$. The integral on the
right-hand side of Eq.~(\ref{eq:the_QI}) therefore converges thereby
providing a finite bound for all $g\in C_0^\infty(\real)$. $\Box$


\section{Examples} \label{sect:examples}

\subsection{Minkowski Spacetime}\label{sect:Mink}

Although Minkowksi space lies outside the discussion of
Sec.~\ref{sect:quantisation} because its Cauchy surfaces are noncompact,
the arguments presented in Sec.~\ref{sect:QWEI} apply equally well to the
two-point functions arising from standard Minkowski quantisations of the Maxwell and
Proca fields based on the Poincar\'e invariant vacua. As we will see,
these vacua are Hadamard; furthermore the topology of the Cauchy surface
played no role in the proof of our QWEIs. We refer the reader
to~\cite{Grundling_Lledo2000} (and references therein) for a careful
discussion of the quantisation of constrained systems in the context of
algebraic field theory in Minkowski space. 

We illustrate our QWEIs for the case of an inertial worldline 
$\gamma$, using the usual Poincar\'e invariant vacua as reference states. 
Boosting to the rest frame of $\gamma$, we adopt coordinates in which 
$\gamma(\tau) = (\tau,x_0,y_0,z_0)$ for some fixed
$(x_0,y_0,z_0)\in\real^3$. We also adopt the orthonormal frame
$v_i^\mu= (\partial/\partial x^i)^\mu$ for $i=0,\dots,3$. 

Let us begin by recalling that the vacuum two-point function for
the {\em scalar} Klein--Gordon equation is given by
\begin{equation}
W_\mass^{(s)}(x,x') = \int_{\real^3}\frac{d^3\bk}{(2\pi)^3} \frac{1}{2\omega} 
e^{-ik_l(x-x')^l}
\end{equation}
where $k^j=(\omega,\bk)$ and $\omega=(|{\bk}|^2+\mass^2)^{1/2}$. 
This is of course a
Hadamard $(\Box+\mass^2)$-bisolution, and it may be used to define
a one-form Hadamard $(\Box+\mass^2)$-bisolution by
\begin{equation}
W_\mass(f,f') = -\eta^{ij}W_\mass^{(s)}(f_i,f_j)\,.
\end{equation}
(The overall sign is determined by the requirement that---anticipating
the commutation relations P4/M4---the
antisymmetric part $W_\mass(f,f')-W_\mass(f',f)$ should be equal
to $-iE_\mass(f,f')=-i\eta^{ij}E^{(s)}_\mass(f_i,f_j)$.)

Turning to the Proca field, the Poincar\'e invariant vacuum two-point
function \cite{Itz&Zu}
\begin{equation}
\omega_0^{(2)}(f,f') = -\int_{\real^3}\frac{d^3\bk}{(2\pi)^3} \frac{1}{2\omega} 
\left( \eta^{ij} - \frac{k^i k^j}{\mass^2} \right)
\widehat{f_i}(-k) \widehat{f_j}(k) 
\end{equation}
is easily seen to be of the form
\begin{equation}
\omega_0^{(2)}(f,f')=W_M(f,(\id-\mass^{-2}\bd\bdelta)f')
\end{equation}
and is therefore Hadamard. We may also write the vacuum two-point function in
terms of its kernel
\begin{equation}
\omega^{(2)ij}_0(x,x') = - \int_{\real^3}\frac{d^3{\bk}}{(2\pi)^3}
\frac{1}{2\omega}\left( \eta^{ij} - \frac{k^i k^j}{\mass^2} \right)
e^{-ik_l(x-x')^l}\,.
\end{equation}
It is not a difficult calculation
to show that the point-split vacuum energy density along the worldline
$\gamma$ is
\begin{equation}
\varphi^*\langle T \rangle_{\omega_0}(\tau, \tau') = \frac{3}{2} 
\int_{\real^3} \frac{d^3{\bk}}{(2\pi)^3}\, \omega e^{-i\omega(\tau-\tau')}
=\frac{6\pi}{(2\pi)^3}\int_0^\infty d\kappa\, \kappa^2 \omega(\kappa) 
e^{-i\omega(\kappa)(\tau-\tau')}\,,
\end{equation}
where, in the last step, we have changed to spherical polar coordinates in the
momentum integration, performed the angular integrals, and written
$\omega(\kappa) = \sqrt{\kappa^2+\mass^2}$.
This expression can be used to evaluate the right-hand side of
Eq.~(\ref{eq:the_QI}), giving
\begin{eqnarray}
\int d\tau\,\langle:\rho:\rangle_\omega(\tau) g(\tau)^2 &\geq&
-\int_0^\infty
\frac{d\alpha}{\pi} \left[ (g\otimes g) \varphi^*\langle T \rangle_{\omega_0}\right]^\wedge
(-\alpha,\alpha)
\nonumber\\
&=& -\frac{6}{(2\pi)^3} \int_0^\infty d\alpha \int_0^\infty
d\kappa\, \kappa^2 \omega(\kappa) \left| 
\widehat{g}\left(\alpha+\omega(\kappa)\right)\right|^2\,.\nonumber
\end{eqnarray}

Before commenting on the above expression, we wish to perform a similar
analysis for the electromagnetic field. The vacuum two-point function
for electromagnetism in the Coulomb gauge is given by
\begin{equation}
\omega^{(2)ij}_0(x,x') = - \int_{\real^3}\frac{d^3{\bk}}{(2\pi)^3}
\frac{1}{2\omega} \,h^{ij}\, e^{-ik_l(x-x')^l},
\end{equation}
where $\omega=|{\bk}|$ and
\begin{equation}
h^{ij} =\left\{ \begin{array}{cl}
\eta^{ij}+k^i k^j/\omega^2 & \hbox{for $i\not=0\not=j$}\\
0 & \hbox{otherwise}.
\end{array}\right.
\end{equation}
It is straightforward to check that, for co-closed test one-forms
$f,f'$ [i.e., $k^i\widehat{f_i}(k)=0$] we have $\omega^{(2)}_0(f,f') =
W_0(f,f')$ and that $\omega_0$ is therefore Hadamard. 
An identical computation to the above now reveals that
\begin{equation}
\varphi^*\langle T \rangle_{\omega_0}(\tau, \tau') = 
\int_{\real^3} \frac{d^3{\bk}}{(2\pi)^3}\, \omega e^{-i\omega(\tau-\tau')}
=\frac{4\pi}{(2\pi)^3}\int_0^\infty d\kappa\, \kappa^2 \omega(\kappa) 
e^{-i\omega(\kappa)(\tau-\tau')}\,,
\end{equation}
and hence
\begin{equation}
\int d\tau\,\langle:\rho:\rangle_\omega(\tau) g(\tau)^2 \geq
-\frac{4}{(2\pi)^3} \int_0^\infty d\alpha \int_0^\infty
d\kappa\, \kappa^2 \omega(\kappa) \left| \widehat{g}\left(\alpha+\omega(\kappa)\right)\right|^2\,.
\end{equation}

{}From this point on, the Proca and electromagnetic fields can be treated
simultaneously as the two bounds take the common form (setting $\mass=0$
for electromagnetism)
\begin{equation}
\int d\tau\,\langle:\rho:\rangle_\omega(\tau) g(\tau)^2 \geq -s I_\mass(g)\,,
\label{eq:comQI}
\end{equation}
where $s$ is the number of spin degrees of freedom for the field, i.e.
$s=2$ for Maxwell and $s=3$ for Proca, and
\begin{equation}
I_{\mass}(g) = \frac{2}{(2\pi)^3} \int_0^\infty
d\alpha \int_0^\infty d\kappa\, \kappa^2 \omega(\kappa) \left| \widehat{g}\left(\alpha+
\omega(\kappa)\right)\right|^2\,.
\end{equation}
This integral can be further simplified by
making an additional change of variables,
\begin{eqnarray}
u&=&\alpha+\omega(\kappa),\nonumber\\
v&=&\omega(\kappa)\,,
\end{eqnarray}
whereupon the above integral can be rewritten as
\begin{equation}
I_{\mass}(g) = \frac{2}{(2\pi)^3} \int_\mass ^\infty
du \left| \widehat{g}(u)\right| ^2 \int_\mass ^u dv \, v^2 \sqrt{v^2-\mass^2}.
\end{equation}
In fact, the scalar field also satisfies Eq.~(\ref{eq:comQI}) with
$s=1$~\cite{Fe&E98}. 
Thus, in Minkowski space, the QWEI bounds for 
scalar, Proca and Maxwell fields differ only by a factor equal to the
number of spin states (as
already noted for the electromagnetic field in~\cite{Pfen02}). However, in more general
curved spacetimes the scalar and one form fields can have different
eigenfrequency spectra leading to very different quantum weak energy inequalities for
each field.

\subsection{Ultrastatic Spacetimes}\label{sec:Static_QI}

We now turn to the class of examples in which the spacetime is ultrastatic
(with compact Cauchy surface obeying our usual requirements) and 
the energy density is sampled along a static worldline, using the
ultrastatic vacua of Sect.~\ref{sect:ultra} as reference
states. In this case, the quantum inequality 
Eq.~(\ref{eq:the_QI}) takes a remarkably simple form. Recall that,
for both the Proca and the Maxwell fields, the two-point
function of the vacuum state can be written compactly as
\begin{equation}
\omega^{(2)}(F,F') = \sum_{\lambda~{\rm physical}}\sum_{j\in J(\lambda)} 
\frac{1}{2\omega(\lambda,j)} 
\langle \A(\lambda ,j) ,F\rangle \,\langle
\overline{\A(\lambda,j)} ,F'\rangle
\label{eq:twopoint_modes}
\end{equation}
where $\lambda$ ranges only over the physical families of modes, i.e.
$\lambda\in\{T,P\}$ for Proca, but $\lambda=T$ only for Maxwell. In
the Maxwell case, we restrict attention to coclosed test one-forms $F$
and $F'$. 
Using the mode expansion~(\ref{eq:twopoint_modes}) for the
two point function, and recalling that $\F=\bd\A$, it is not difficult
to show that \footnote{Effectively, one is substituting the
test function $f(x) = \int d\tau |h(x)|^{-1/2} \delta^4(x-\gamma(\tau))
\overline{g_\alpha(\tau)}$ in Eq.~(\ref{eq:smeared_unrenorm_T}). This results in the required expression.}
\begin{equation}
\left[(g \otimes g)\varphi^*\langle T\rangle_\omega \right]^\wedge (-\alpha,\alpha)
=\int d\tau \int d\tau' g(\tau)\, g(\tau')\,e^{-i\alpha(\tau-\tau')}
\sum_{\lambda~{\rm physical}} \sum_{j\in J(\lambda)} \, N(\lambda,j)(\gamma(\tau),\gamma(\tau'))\,,
\label{eq:argument_part}
\end{equation}
where
\begin{eqnarray}
N(\lambda,j)(x,x') &=& \frac{1}{2\omega(\lambda,j)}\left\{ \frac{1}{4}
\sum_{p=0}^3\sum_{q=0}^3 \left[ \F_{\mu\nu}(\lambda,j) v_p^\mu
v_q^\nu\right](x)\,\left[ \overline{\F_{\rho\sigma}(\lambda,j)} v_p^\rho
v_q^\sigma\right](x')\right.\nonumber\\
&&\left. +\frac{1}{2}\mass^2 \sum_{p=0}^3
\left[ \A_{\mu}(\lambda,j) v_p^\mu \right](x)\,\left[ \overline{
\A_{\rho}(\lambda,j)} v_p^\rho\right](x')\right\}\,.
\end{eqnarray}
Now if we choose a static worldline $\gamma(\tau)=(\tau, {\bf x}_0)$,
then
\begin{equation}
N(\lambda,j)(\gamma(\tau),\gamma(\tau'))=e^{-i\omega(\lambda,j)(\tau-\tau')}
\,\fT(\lambda,j)\,,
\end{equation}
where
\begin{equation}
\fT(\lambda,j)
= \frac{1}{2\omega(\lambda,j)}\left\{ \frac{1}{4}
\sum_{p=0}^3\sum_{q=0}^3 \left|\left[ \F_{\mu\nu}(\lambda,j) v_p^\mu
v_q^\nu\right](\gamma(\tau_0))\right|^2
 +\frac{1}{2}\mass^2 \sum_{p=0}^3\left|
\left[ \A_{\mu}(\lambda,j) v_p^\mu \right](\gamma(\tau_0))\right|^2\right\}
\label{eq:class_E_density}
\end{equation}
is the classical energy density per mode at the spatial position of the
observer's worldline. (Note that the leading factor of
$(2\omega(\lambda,j))^{-1}$ appears because our modes were normalised
using the $L^2$-inner product on $\Sigma$, rather than the symplectic
inner product.) The most notable point to be made here is that
$\fT(\lambda,j)$
is $\tau$ independent because the $e^{-i\omega_\lambda \tau}$
dependence in the modes is removed by taking the magnitude of the
complex functions. Thus eq.~(\ref{eq:argument_part}) becomes
\begin{equation}
\left[(g \otimes g)\varphi^*\langle T\rangle_\omega \right]^\wedge
(-\alpha,\alpha)
= \sum_{\lambda~{\rm physical}} \sum_{j\in J(\lambda)} \left|\widehat{g}(\alpha+\omega(\lambda,j))
\right|^2 \fT(\lambda,j)\,,
\end{equation}
and the quantum weak energy inequality for static observers in 
ultra-static spacetimes becomes
\begin{equation}
\int d\tau (g(\tau))^2 \langle:\rho:\rangle_\omega(\tau)
\geq -\int_0^\infty \frac{d\alpha}{\pi}
\sum_{\lambda~{\rm physical}} \sum_{j\in J(\lambda)} \left|\widehat{g}(\alpha+\omega(\lambda,j))
\right|^2 \fT(\lambda,j). 
\label{eq:the_ultra_static_QI}
\end{equation}
As we will see, the above form turns out to be calculationally convenient in some
settings. However, we also note
that the quantum weak energy inequality may be written
\begin{equation}
\int d\tau (g(\tau))^2 \langle:\rho:\rangle_\omega(\tau)
\geq -\int_0^\infty du\, |\widehat{g}(u)|^2 Q(u)
\end{equation}
where 
\begin{equation}
Q(u) = \frac{1}{\pi} \sum_{\lambda~{\rm physical}} \sum_{\stack{j\in
J(\lambda)}{{\rm s.t.}~\omega(\lambda,j)\le u}} \fT(\lambda,j)
\end{equation}
measures the maximum energy density available if no mode with frequency
greater than $u$ is excited. The analysis carried out in Sect.~5 of~\cite{Fews00} for the 
scalar field may be adapted to show that $Q(u)$ is positive and 
polynomially bounded as $u\to\infty$. Several other systems obey quantum inequalities 
of this form, which motivated the study of related conditions for
general quantum dynamical systems in~\cite{Fe&V03}. In 
the terminology of that paper, the ultrastatic vacuum states 
constructed here fulfill a {\em limiting static QWEI}. We expect that the
analysis of~\cite{Fe&V03} would apply equally well to the Proca and
Maxwell fields, thereby establishing links between quantum inequalities
and the thermodynamic properties of these fields.

\subsection{Static Einstein Universe}
\label{sect:Einstein}

A nice example of the ultrastatic quantum inequality is provided by the static
Einstein universe $\real\times S^3$ with line element
\begin{equation}
ds^2=dt^2-a^2\left[ d\chi^2 + \sin^2\chi\left(d\theta^2+\sin^2\theta
d\varphi^2\right)\right].
\end{equation}
Here $a$ is the radius of the universe and $(\chi,\theta,\varphi)$ are
spherical polar coordinates on the unit three-sphere. 
Various authors have studied the electromagnetic mode functions in the Einstein universe
\cite{Lifs46,Park72,Mash73} and
it turns out that essentially the same ansatz can be used to obtain
the modes of the Proca field.

We know from Section~\ref{sect:ultra} that the scalar Proca modes 
are completely specified by finding mode solutions to the
massive, minimally-coupled scalar wave equation
$\left(\Box^s +\mass^2\right)\Phi=0$. A complete set of suitably normalised
positive frequency mode solutions in the Einstein universe is given by 
\cite{Parker,Ford76}
\begin{equation}
\Phi_{nlm}(x) = a^{-3/2} \Pi_{nl}(\chi)\, Y_{lm}(\theta,\varphi)\,e^{-i\sigma_n t},
\end{equation}
where $Y_{lm}(\theta,\varphi)$ are the standard spherical harmonics on
a two-sphere \cite{Jackson}, $\Pi_{nl}(\chi)$ can be written in
terms of Gegenbauer polynomials $C^\lambda_\eta(x)$ \cite{Gradshteyn} as 
\begin{equation}
\Pi_{nl}(\chi)=\frac{l!\, 2^{l+\frac{1}{2}}\sqrt{(n-l)!(n+1)}}{\sqrt{\pi(n+l+1)!}}
\sin^{l}\chi\, C^{l+1}_{n-l}(\cos\chi),
\end{equation}
and the frequency of the modes is
\begin{equation}
\sigma_n = \sqrt{\frac{n(n+2)}{a^2}+\mass^2}.
\label{eq:freq_axial}
\end{equation}
Here, the primary quantum number $n$ ranges
over the non-negative integers $n=0,1,2,\dots$. For a given $n$ there are $(n+1)^2$
harmonic states with the same frequency labelled by the quantum numbers,
$l=0,1,\dots,n$ and $m=-l,{-l+1},\dots,0,\dots,l-1,l$. However, the case
$n=l=m=0$ corresponds to the spatially constant mode; thus the labelling
set for the scalar Proca modes given by Eq.~(\ref{eq:scalar_proca_def})
is $n=1,2,\ldots$, with $l$ and $m$ as before. 

On the other hand, the family of transverse modes breaks up
into two subfamilies \cite{Mash73}: the magnetic J-pole modes taking the form
\begin{equation}
\A(M,n,l,m) = *\bd \left(\Psi_{nlm}\bd t\wedge\bd\chi\right)\,,
\end{equation}
and the electric J-pole modes, 
\begin{equation}
\A(E,n,l,m) = *\bd \left(\frac{a}{n}\A(M,n,l,m)\wedge \bd t\right)\,,
\end{equation}
where the scalar functions $\Psi_{nlm}$
obey
\begin{equation}
\left(\Box^s + \frac{2\cos\chi}{a^2\sin\chi}
\partial_\chi + \mass^2\right)\Psi_{nlm}=0.
\label{eq:Einstein_PDE}
\end{equation}
The positive frequency mode solutions to this partial differential
equation are
\begin{equation}
\Psi_{nlm}(x) = \sqrt{2a n} \, V_{nl}(\chi)\, Y_{lm}(\theta,\varphi)\, 
e^{-i\omega_n t},
\label{eq:Psi_modes}
\end{equation}
where
\begin{equation}
V_{nl}(\chi)=\frac{l!\, 2^l \sqrt{(n-l-1)!}}{\sqrt{\pi l(l+1)(n+l)!}}
\sin^{l+1}\chi\, C^{l+1}_{n-l-1}(\cos\chi)
\end{equation}
and the frequency of the modes is given by
\begin{equation}
\omega_n = \sqrt{\frac{n^2}{a^2}+\mass^2}.
\label{eq:freq_trans}
\end{equation}
Here the quantum numbers range over a different set of allowed values.
The primary quantum number $n$ ranges over the integers $n=2,3,\dots$. 
For a given $n$ there are $n^2-1$ harmonic states
with the same frequency labelled by the quantum numbers, $l=1,2,\dots,n-1$ and
$m=-l,-1+1,\dots,0,\dots,l-1,l$.  We stress that
both transverse sub-families have an energy spectrum given by 
Eq.~(\ref{eq:freq_trans}), which differs from that of the scalar Proca
modes. There is actually a more convenient basis of transverse mode
solutions given by
\begin{equation}
\A(\pm ,n,l,m)= \frac{1}{\sqrt{2}} \left[\A(M,n,l,m) \pm \A(E,n,l,m)\right]\,,
\end{equation}
which are simultaneous eigenfunctions of $\bdelta_\Sigma
*_\Sigma$ and $-\Delta_\Sigma +\mass^2$ (which commute). The $\lambda=M$ or $E$ families of
modes are, roughly speaking, the linearly polarized solutions of the field
while the $\lambda = +$ or $-$ modes are the circularly polarized solutions.

For the interested reader, we also state the component form of the three families of
physical modes.
The two transverse polarizations of the field are
\begin{equation}
\A_\mu(M,n,l,m) =\frac{1}{a}\left( 0,\, 0,\, -\frac{1}{\sin\theta}
\partial_\varphi,\, \sin\theta \partial_\theta \right)\Psi_{nlm},
\label{eq:magnetic_Jpole}
\end{equation}
and 
\begin{equation}
\A_\mu(E,n,l,m) = \frac{1}{ a n}\left( 0,\, \frac{l(l+1)}
{\sin^2\chi},\, \partial_\chi\partial_\theta,\, \partial_\chi
\partial_\varphi\right) \Psi_{nlm}\,,
\label{eq:electric_Jpole}
\end{equation}
while the scalar Proca modes are
\begin{equation}
\A_\mu(P,n,l,m) = \frac{a \sigma_n^2}{M \sqrt{n (n+2)}}\left(
\frac{n(n+2)}{i\sigma_n a^2},\, \partial_\chi,\, \partial_\theta,\, 
\partial_\varphi\right) \Phi_{nlm}\,.
\label{eq:Axial}
\end{equation}
One may readily verify that the two transverse modes and the 
scalar Proca mode all have the property $\bdelta A(\lambda,n,l,m)=0$,
for $\lambda=M,E,P$ and therefore solve the Proca 
equation~(\ref{eq:proca_wave_eq}). Moreover, these modes
obey the (pseudo)-orthonormality conditions of Sect.~\ref{sect:constr}.

Since the Einstein spacetime is ultra-static, we can find the quantum weak
energy inequality from Eq.~(\ref{eq:the_ultra_static_QI}).  To begin, 
let the worldline for a stationary observer be given by $\gamma(\tau)=
(\tau,\chi,\theta,\varphi)$ where $(\chi,\theta,\varphi)$ are
the coordinates of a fixed point in space. An orthonormal basis in the
neighbourhood around $\gamma$ is given by $v_i^\mu = \sqrt{|g^{\mu\mu}|}
(\partial/\partial x^i)^\mu$ [no sum on $\mu$].  All of the work
is in the evaluation the classical energy density per mode, $\fT(\lambda,n,l,m)$,
defined by Eq.~(\ref{eq:class_E_density}), which, in the Einstein spacetime
takes the form
\begin{eqnarray}
\fT(\lambda,j) &=& \frac{1}{4\omega(\lambda,j)} \left[ 
  \frac{\left|\F_{01} \right|^2}{a^2}  
+ \frac{\left|\F_{02} \right|^2}{a^2 \sin^2\chi}  
+ \frac{\left|\F_{03} \right|^2}{a^2 \sin^2\chi \sin^2\theta} 
+ \frac{\left|\F_{12} \right|^2}{a^4 \sin^2\chi}  
+ \frac{\left|\F_{13} \right|^2}{a^4 \sin^2\chi \sin^2\theta}  
+ \frac{\left|\F_{23} \right|^2}{a^4 \sin^4\chi \sin^2\theta}
\right.\nonumber\\ 
&&\left. +\mass^2 \left( \left|\A_{0} \right|^2
+ \frac{\left|\A_{1} \right|^2}{a^2}  
+ \frac{\left|\A_{2} \right|^2}{a^2 \sin^2\chi}  
+ \frac{\left|\A_{3} \right|^2}{a^2 \sin^2\chi \sin^2\theta} 
\right)\right].
\end{eqnarray}

Beginning with the $\pm$ modes, it is straightforward to calculate the
field strength $\F(\pm,n,l,m)=\bd\A(\pm,n,l,m)$. Inserting the field
strength into the above expression, and using  $\partial_\chi
\Psi_{nlm}=\partial_\chi \overline{\Psi_{nlm}}$, we arrive at 
\begin{eqnarray}
\fT(\pm,n,l,m) &=& \frac{\omega_n}{4 a^4 n^2} \left[ \left( \frac{l(l+1)}
{\sin^2\chi}\right)^2 \left|\Psi_{nlm}\right|^2 + \frac{1}{\sin^2\chi}
\left( \left|\partial_\chi \partial_\theta \Psi_{nlm}\right|^2+
\frac{1}{\sin^2\theta}\left|\partial_\chi \partial_\varphi\Psi_{nlm}
\right|^2  \right) \right.\nonumber\\
&& \left. + \frac{n^2}{\sin^2\chi}
\left( \left| \partial_\theta \Psi_{nlm}\right|^2+
\frac{1}{\sin^2\theta}\left| \partial_\varphi\Psi_{nlm}
\right|^2  \right)\right]\,.
\end{eqnarray}
Note that the $+$ and $-$ modes have equal energy densities.
Next, we use the definition Eq.~(\ref{eq:Psi_modes}) of the scalar
functions $\Psi_{nlm}$ and two identities. The first identity,
\begin{equation}
\left|\partial_\theta Y_{lm}\right|^2+
\frac{1}{\sin^2\theta}\left|\partial_\varphi Y_{lm}\right|^2
= \frac{1}{2} \Delta_{S^2} \left|Y_{lm}\right|^2+l(l+1)\left|Y_{lm}\right|^2
\end{equation}
is for the spherical harmonics where
\begin{equation}
\Delta_{S^2}= \frac{1}{\sin\theta}\partial_\theta \sin\theta \partial_\theta
+\frac{1}{\sin^2 \theta}\partial_\phi^2
\end{equation}
is the two-sphere Laplacian. The second identity is
\begin{equation}
\frac{l(l+1)}{\sin^2\chi}V_{nl}^2+\left(\partial_\chi V_{nl}\right)^2=
\left(n^2+\frac{1}{2}\partial_\chi^2 \right) V_{nl}^2\,,
\end{equation}
which follows from the ordinary differential equation satisfied by $V_{nl}(\chi)$.
Using both of these we arrive at
\begin{equation}
\fT(\pm,n,l,m) = \frac{\omega_n}{2 a^3 n} \left\{  \frac{l(l+1)}
{\sin^2\chi} \left(2n^2+\frac{1}{2}\partial_\chi^2 \right) V_{nl}^2
\left| Y_{lm} \right|^2
+ \frac{1}{2\sin^2\chi}\left[ \left( \partial_\chi V_{nl}\right)^2
+ n^2 V_{nl}^2\right] \Delta_{S^2} \left| Y_{lm} \right|^2
\right\}.
\end{equation}

Because the frequency $\omega_n$ is independent of $l$ and $m$, we can
sum over these labels, obtaining considerable simplification
from the summation theorems for 
the spherical harmonics and Gegenbauer polynomials. 
In the case of the spherical harmonics it is well-known
that \cite{Jackson}
\begin{equation}
\sum_{m=-l}^l \left| Y_{lm}(\theta,\varphi)\right|^2 =\frac{2l+1}{4\pi}\,, 
\end{equation}
which reduces the sum over all $l$ and $m$ of $\fT(\pm,n,l,m)$ to
\begin{eqnarray}
\sum_{l=1}^{n-1}\sum_{m=-l}^{+l} \T(\pm,n,l,m) &=& \frac{\omega_n}{8 \pi a^3 n}
\sum_{l=1}^{n-1}\frac{(2l+1)l(l+1)}
{\sin^2\chi} \left(2n^2+\frac{1}{2}\partial_\chi^2 \right) V_{nl}^2\,,\nonumber\\
&=&\frac{\omega}{8 \pi a^3 n}\frac{1}{\sin^2\chi}\left(2n^2+\frac{1}{2}
\partial_\chi^2 \right) \sin^2\chi \sum_{l=1}^{n-1}\frac{(2l+1)l(l+1)}
{\sin^2\chi}V_{nl}^2.
\end{eqnarray}
Using the summation theorem for the Gegenbauer polynomials, 
\begin{equation}
\sum_{l=1}^{n-1} \frac{(2l+1)l(l+1)}{\sin^2\chi}V_{nl}^2(\chi) = \frac{1}
{\pi}\left[ n- \frac{1}{n}\left( \frac{\sin n\chi}{\sin\chi}\right)^2\right]\,,
\end{equation}
[Eq.~8.934.3 of \cite{Gradshteyn}] we finally obtain
\begin{equation}
\sum_{l=1}^{n-1}\sum_{m=-l}^{+l} \fT(\pm,n,l,m) = \frac{(n^2-1)\omega_n}{4 \pi^2 a^3}\,.
\label{eq:transverse_E}
\end{equation}
It is noteworthy that this expression is exactly the total zero-point
energy of $n^2-1$ harmonic oscillators at frequency $\omega_n$, divided
by the spatial volume $2\pi^2 a^3$ of the
Einstein universe. 

The contribution from the scalar Proca modes is calculated in much the same manner.
The only difference is that we make use of the following two identities, 
\begin{equation}
\left(\partial_\chi \Pi_{nl} \right)^2 +\frac{l(l+1)}{\sin^2\chi}
\Pi_{nl}^2 = n(n+2) \Pi_{nl}^2+ \frac{1}{2\sin^2\chi}\partial_\chi
\sin^2\chi\partial_\chi \Pi_{nl}^2\,,
\end{equation}
which follows from the ordinary differential equation satisfied
by $\Pi_{nl}$ and
\begin{equation}
\sum_{l=0}^n (2l+1)\Pi_{nl}^2(\chi) = \frac{2(n+1)^2}{\pi}\,,
\end{equation}
which again follows from the summation theorem for the Gegenbauer
polynomials.  For the Proca mode we find
\begin{equation}
\sum_{l=0}^{n}\sum_{m=-l}^{+l} \fT(P,n,l,m) = \frac{(n+1)^2\sigma_n}{4 \pi^2 a^3}.
\label{eq:Proca_E}
\end{equation}
Again we have found the zero-point energy times the multiplicity divided by the
spatial volume.

Finally, let $s$ be the total number of spin degrees of freedom for the field,
i.e. $s=2$ for Maxwell and $s=3$ for Proca.  Substituting
Eqs.~(\ref{eq:transverse_E}) and (\ref{eq:Proca_E})
into Eq.~(\ref{eq:the_ultra_static_QI}), we can write the quantum weak energy
inequality in the Einstein spacetime for an arbitrary real-valued test function
$g(\tau)\in C^\infty_0(\real)$ as
\begin{equation}
\int d\tau\,\langle:\rho:\rangle_\omega(\tau) g(\tau)^2 \geq
-\frac{2}{4\pi^2 a^3}\sum_{n=2}^\infty(n^2-1)\omega_n 
\int_0^\infty \frac{d\alpha}{\pi}
\left| \widehat{g}(\alpha+\omega_n)\right|^2 
- \frac{(s-2)}{4\pi^2 a^3}\sum_{n=1}^\infty (n+1)^2 \sigma_n
\int_0^\infty \frac{d\alpha}{\pi}
\left| \widehat{g}(\alpha+\sigma_n)\right|^2  \,,
\label{eq:Einstein_QWEI}
\end{equation}
which holds for all Hadamard states $\omega$. 

\begin{figure}
\rotatebox{0}{\scalebox{1.00}{
\includegraphics{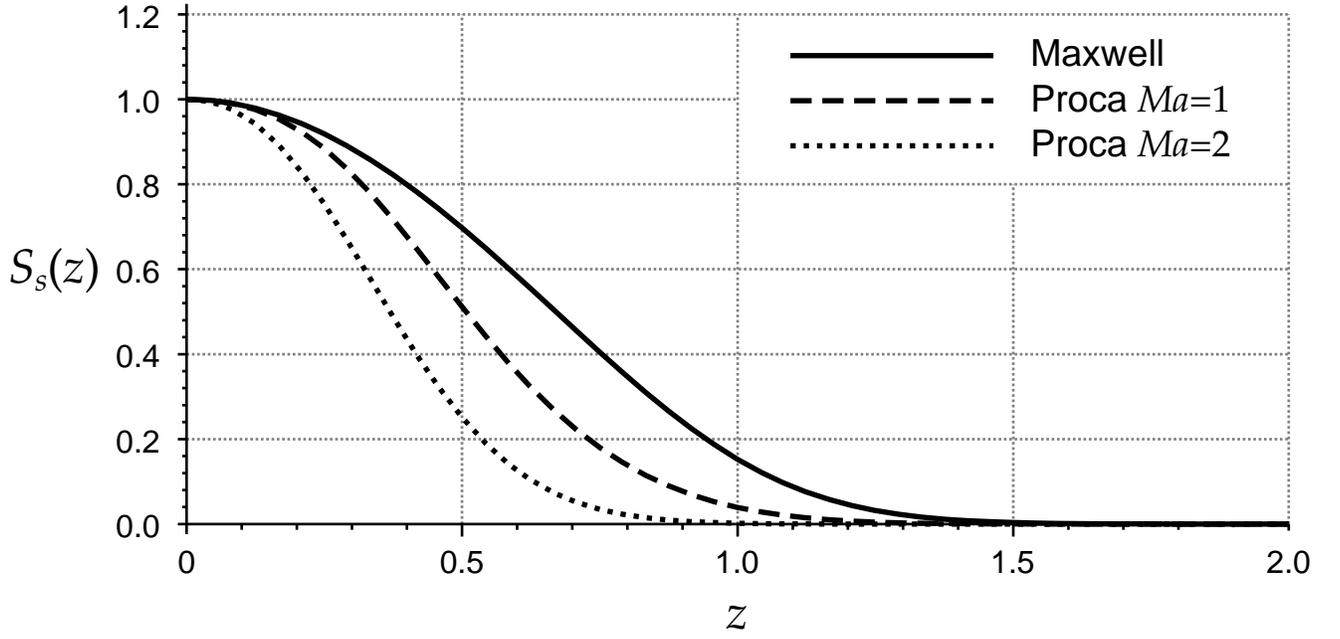} 
}}
\caption{\label{fig:Einstein_gaussian} Plot of the scale function $\S_s(z)$
for the Gaussian sampling function in the Einstein Universe. Three
different cases are of Eq.~(\ref{eq:Einstein_S}) are displayed: 
Electromagnetism (solid line) and two cases of the Proca field for
different values of the mass times the radius of universe, $\mass a= 1$ 
(dashed line) and $\mass a= 2$ (dotted line).   
The plots were produced by evaluating the first 3000 terms in the summation
of Eq.~(\ref{eq:Einstein_S}).}
\end{figure}

By suitably restricting the class
of allowed states (if necessary) this bound may be extended to
noncompactly supported $g$ whose Fourier transforms have sufficiently
rapid decay~\footnote{The RHS of the bound may be written in the form
$-\int_0^\infty du\, Q(u) |\widehat{g}(u)|^2$, where $Q(u)$ is a
polynomially bounded function. Accordingly, the integral will converge
provided that the Fourier transform $\widehat{g}(u)$ decays sufficiently
rapidly as $u\to+\infty$; equivalently, provided that $g$ belongs to a
Sobolev space $W^{k,2}(\real)$ (see e.g.,~\cite{Adams75}) of sufficiently high
regularity ($k$ sufficiently large). If we restrict the class
of states so that $\langle:\rho:\rangle_\omega(\tau)$ belongs to the
dual Sobolev space $W^{-k,2}(\real)$ then Eq.~(\ref{eq:Einstein_QWEI}) will continue to
hold. To see this, we approximate $g$ in the $W^{k,2}(\real)$ norm by a sequence of compactly
supported smooth functions [using density of $C_0^\infty(\real)$ in
$W^{k,2}(\real)$] and apply Eq.~(\ref{eq:Einstein_QWEI}). It is
not presently known whether this requires a nontrivial restriction on
the class of allowed states.}.
For electromagnetism, the above bound was evaluated for a Lorentzian test function
in~\cite{Pfen02}; here we will treat the example of a Gaussian test function in
both the electromagnetic and Proca cases. Setting
\begin{equation}
g(\tau) = \pi^{-1/4}\tau_0^{-1/2} \exp\left(-\frac{1}{2}
\frac{\tau^2}{\tau_0^2}\right)\,,
\end{equation}
the weight function $g(\tau)^2$ is a normalised Gaussian. Using
the Fourier transform of $g$,
\begin{equation}
\widehat{g}(\alpha)= \pi^{1/4} (2\tau_0)^{1/2} \exp\left(-\frac{\alpha^2\tau_0^2}{2}\right)\,,
\end{equation}
it is now possible to calculate
\begin{equation}
\int_0^\infty d\alpha \left|\widehat{g}(\alpha+\omega)\right|^2 = \pi\,
\mbox{erfc} (\omega\tau_0)\,,
\end{equation}
where $\mbox{erfc}(x)$ is the complementary error function. Accordingly, the quantum
weak energy inequality for this test function is
\begin{equation}
\int d\tau\,\langle:\rho:\rangle_\omega(\tau) g(\tau)^2 \geq -\frac{3s}
{64\pi^2\tau_0^4}\,\S_s(\tau_0/a),
\end{equation}
where the scale function
\begin{widetext}
\begin{eqnarray}
\S_s(z)&=& \frac{16}{3} z^4 \left[ \frac{2}{s} \sum_{n=2}^\infty (n^2-1)
\sqrt{n^2+\mass^2a^2}\,\mbox{erfc} (\sqrt{n^2+\mass^2a^2}z)\right.\nonumber\\
&&\left.+\frac{s-2}{s}\sum_{n=1}^\infty (n+1)^2\sqrt{n(n+2)+\mass^2a^2}\,
\mbox{erfc}(\sqrt{n(n+2)+\mass^2a^2}z) \right]
\label{eq:Einstein_S}
\end{eqnarray}
\end{widetext}
is plotted in Figure~\ref{fig:Einstein_gaussian}.
For very short sampling times the scale function is close to unity 
and the QWEI bound is the same as that for Minkowski
spacetime \cite{Pfen02}. This makes sense because the spacetime appears to be flat over time scales which
are short relative to the radius of the universe, and
we should expect to recover the Minkowski space result.
As the sampling time becomes progressively longer, the field
has more time to ``sample'' the curvature of the universe, and thus it
becomes increasingly difficult to generate negative energy densities.
Also, as the mass of the field increases it becomes increasingly 
difficult to generate negative energy densities.  This is seen by the faster decay
in the scale functions for larger value of the field mass as
seen in previous evaluations of the scale function for scalar fields in other spacetimes
\cite{F&Ro97,Pfen98a,Pfen98b}.  As mentioned above, the quantum
inequality bound has also been evaluated for electromagnetism with
a Lorentzian sampling function \cite{Pfen02} and the
behaviour is very much the same.

\appendix

\section{Consistency of the Hadamard condition with the constraints}
\label{appx:cons}

In this appendix we prove the following local-to-global result.
\begin{Theorem} \label{thm:con}
Let $\omega$ be a state on either $\mathfrak{A}_\mass (\M,g)$ or
$\mathfrak{A}(\M,g)$ and suppose that there exists a causal normal neighbourhood
$\N$ of a Cauchy surface $\Sigma$ in $\M$, and a Hadamard bisolution $W_\mass$ (respectively, $W$) 
to $\Box+\mass^2$ (respectively, $\Box$) on $\N\times\N$ such that
\begin{equation}
\omega\left( \A (f_1) \A (f_2) \right)
= W_\mass\left(f_1,\left(\hbox{{\rm $\id$}} - \mass^{-2}\bd\bdelta\right)f_2\right) 
\label{eq:locHad_Proca}
\end{equation}
for all $f_i \in \Omega^1_0(\N)$ (in the Proca case) or
\begin{equation}
\omega\left( [\A ](f_1) [\A ](f_2) \right)
= W\left(f_1,     f_2 \right) 
\label{eq:locHad_em}
\end{equation}
for all $f_i \in \Omega^1_0(\N)$ with $\bdelta f_i = 0$ (in the Maxwell case). 
Then $\omega$ is Hadamard. 
\end{Theorem}
Thus the Hadamard conditions introduced in Sect.~\ref{sect:QHad} 
respect the Cauchy evolution of the field equations: if they are
satisfied near one Cauchy surface, they hold near all others and
hence hold globally. 
As the Hadamard form for Klein--Gordon bisolutions also propagates in this
sense, our discussion may be regarded as checking the consistency of
the definitions of Sect.~\ref{sect:QHad} with the Proca and Maxwell
constraints. The propagation property is one of the key ingredients 
used in Sect.~\ref{sect:FNW}, where we established 
the existence of Hadamard states on general globally
hyperbolic spacetimes obeying our usual topological conditions. Although
we will use the fact that $\Sigma$ is compact for both the Maxwell and
Proca fields this condition can probably be dropped; however in the
Maxwell case it appears to be necessary [in Lemma~\ref{lem:glob_lem}(b)] 
to assume that $H_1(\Sigma)$ 
is trivial and that the compact support
cohomology group $H^3_c(\M)$ of the spacetime is therefore also trivial.

The proof of Theorem~\ref{thm:con} relies on a number of results concerning the
classical Proca, Maxwell and Klein--Gordon fields. One fact which will
be used repeatedly is that, for $f\in\Omega_0^1(\M)$, $E_Mf=0$ if and
only if $f\in(\Box+\mass^2)\Omega_0^1(\M)$. Indeed,
$f=(\Box+\mass^2)E_M^+f$ and since, by hypothesis, $E_M^+f=E_M^-f$, the
support of $E_M^+f$ is contained in the compact set $J^+(\supp f)\cap J^-(\supp
f)$. Our first observation generalises this fact to the Proca field. 

\begin{Lemma} \label{lem:DMker}
Let $M>0$. For $f\in\Omega_0^1(\M)$, $\Delta_M f=0$ if and
only if $f\in(-\bdelta\bd+\mass^2)\Omega_0^1(\M)$. 
\end{Lemma}
\proof Since $(\id-\mass^{-2}\bd\bdelta)(-\bdelta\bd+\mass^2) =
\Box+\mass^2$, sufficiency holds because $E_M$ is a Klein--Gordon
bisolution. Conversely, $\Delta_Mf=0$ is equivalent to $E_\mass (\id-\mass^{-2}\bd\bdelta)f
=0$; hence, we have
\begin{equation}
(-\bd\bdelta+\mass^2)f =\mass^2(\Box+\mass^2)g
\end{equation}
for some $g\in\Omega_0^1(\M)$. By applying $\bdelta\bd$ to both sides,
we obtain $\bdelta\bd f = (\Box+\mass^2)\bdelta\bd g$, which 
may be subtracted from the previous expression to yield
\begin{equation}
(\Box +\mass^2)f = (\Box+\mass^2)(-\bdelta\bd+\mass^2)g\,.
\end{equation}
Since both $f$ and $g$ are compactly supported, we conclude that
$f=(-\bdelta\bd+\mass^2)g$. $\Box$

\begin{Proposition} \label{lem:eom_test}
(a)~Given any $f\in\Omega_0^1(\M)$ there exists $g\in\Omega_0^1(\M)$ 
such that $\widetilde{f}= f+ (-\bdelta\bd +\mass^2)g$
is an element of $\Omega_0^1(\N)$. \\
(b)~Given any co-closed $f\in\Omega_0^1(\M)$ there exists $g\in\Omega_0^1(\M)$ 
such that $\widetilde{f}= f- \bdelta\bd g$ is a co-closed element of $\Omega_0^1(\N)$.
\end{Proposition}
\proof (a) Choose smooth
functions $\chi^\pm$ on $\M$ with $\chi^++\chi^-=1$ and $\chi^+$ equal to
unity to the future of $\N$ and vanishing to the past of $\N$. Then
\begin{equation}
\widetilde{f}= (-\bdelta\bd +\mass^2) \chi^+ \Delta_\mass f
\end{equation}
belongs to $\Omega_0^1(\N)$ and satisfies 
$\Delta_\mass \widetilde{f} = \Delta_\mass f$. Applying
Lemma~\ref{lem:DMker} to $\widetilde{f}-f$, it follows that
$\widetilde{f}=f+(-\bdelta\bd+\mass^2)g$ as required. \\
(b)~With $\chi^\pm$ as above, set $\widetilde{f}=-\bdelta\bd\chi^+E_0f$.
Then $\widetilde{f}$ belongs to $\Omega_0^1(\N)$ and is co-closed;
moreover $E_0\widetilde{f}$ and $E_0 f$ are gauge equivalent by 
the proof of Prop.~4(c) in~\cite{Dimock}. Accordingly, there exists
$\eta\in\Omega^0(\M)$ such that $E_0(\widetilde{f}-f)=\bd\eta$. Taking
co-derivatives, using $\bdelta E_0=E_0\bdelta$ and co-closure of
$\widetilde{f}$ and $f$, we have $\bdelta\bd\eta=0$ and hence
$\eta=E_0\zeta$ for some $\zeta\in\Omega_0^0(\M)$. Substituting back, we
see that $E_0(\widetilde{f}-f-\bd\zeta)=0$, so
\begin{equation}
\widetilde{f}-f-\bd\zeta=\Box g \label{eq:ffzg}
\end{equation}
for some $g\in\Omega_0^1(\M)$. Taking co-derivatives again, 
$\Box(\zeta-\bdelta g) = 0$ and hence (because both $\zeta$ and $g$ are
compactly supported) $\zeta=\bdelta g$. Substituting back in
Eq.~(\ref{eq:ffzg}), we have $\widetilde{f}=f+\bd\bdelta g+\Box g=
f-\bdelta\bd g$ as required. $\Box$

\begin{Lemma} \label{lem:glob_lem}
(a) If $\A$ is a weak one-form
$(\Box+\mass^2)$-solution obeying
\begin{equation}
\A((-\bdelta\bd+\mass^2)f)=0 \label{eq:loc_proc}
\end{equation}
for all $f\in\Omega_0^1(\N)$, then Eq.~(\ref{eq:loc_proc}) holds for
all $f\in\Omega_0^1(\M)$. \\
(b) If $\A$ is a weak one-form $\Box$-solution vanishing on
co-closed $f\in\Omega_0^1(\N)$, then $\A$ vanishes on all co-closed
$f\in\Omega_0^1(\M)$. 
\end{Lemma}
\proof (a) Since, for $f\in\Omega_0^1(\N)$,
$\A((\Box+\mass^2)f)=0$, we have by Eq.~(\ref{eq:loc_proc}) that
\begin{equation}
\A(\bd\bdelta f)=0 
\label{eq:A8}
\end{equation}
for all $f\in\Omega_0^1(\N)$. Thus, $\A(\bd\cdot)$ is a 
global weak scalar $(\Box+\mass^2)$-solution vanishing on
$\bdelta\Omega_0^1(\N)$. Now we may fix $f_0\in\Omega_0^0(\N)$ such that
any $f\in\Omega_0^0(\N)$ may be written 
\begin{equation}
f = f_0 \int f \,d{\rm vol}_g + \bdelta h
\label{eq:A9}
\end{equation}
for some $h\in\Omega_0^1(\N)$. (Since $\N$ is connected, boundaryless
and orientable, this follows by de~Rahm's theorem:
see the remarks following Theorem~7.5.19 in~\cite{AMR88}.) 
Combining Eqs.~(\ref{eq:A8} and~(\ref{eq:A9}), we have 
$\A(\bd f) = \A(\bd f_0)\int f\,d{\rm vol}_g$ for all
$f\in\Omega_0^0(\N)$. Accordingly,
$\A(\bd \cdot)$ is constant on $\N$ and hence (since it is a
$(\Box+\mass^2)$-solution) on $\M$. It follows that
$\A(\bd\bdelta f) = \A(\bd f_0)\int \bdelta f\,d{\rm vol}_g=0$ for
all $f\in\Omega_0^1(\M)$. Since $\A((\Box+\mass^2)f)=0$, we deduce that
Eq.~(\ref{eq:loc_proc}) holds for all $f\in\Omega_0^1(\M)$ as required.
\\
(b) $\A(\bdelta\cdot)$ is a two-form weak
$\Box$-solution vanishing on all $h\in\Omega_0^2(\N)$ and
hence on all $h\in\Omega_0^2(\M)$. But since $H^3_c(\M)$ is trivial, 
any $f\in\Omega_0^1(\M)$ such that $\bdelta f=0$ can
be written as $f=\bdelta h$ for some $h\in\Omega_0^2(\M)$ and we 
conclude that $\A(f)=\A(\bdelta h) =0$. $\Box$

After these preliminaries, we may now prove the main result of this section.\\
{\noindent\em Proof of Theorem~\ref{thm:con}.} The arguments for the two
theories run along parallel
lines. First one extends the $W_\mass$ (resp., $W$) to be a global bisolution to
the appropriate one-form Klein--Gordon equation. Since the Hadamard form
propagates, it suffices to show that Eq.~(\ref{eq:locHad_Proca}) holds
for all $f_i \in \Omega^1_0(\M)$ (resp., that Eq.~(\ref{eq:locHad_em}) holds
for all co-closed $f_i \in \Omega^1_0(\M)$). In the Proca case, we apply
Proposition~\ref{lem:eom_test}(a) together with the field equation axiom P3
and Eq.~(\ref{eq:locHad_Proca}) 
to show that
\begin{equation}
\omega\left( \A (f_1) \A (f_2) \right) =
\omega\left( \A (\widetilde{f_1}) \A (\widetilde{f_2}) \right)
=W_\mass(\widetilde{f_1},\left(\hbox{{\rm $\id$}} - \mass^{-2}\bd\bdelta\right)\widetilde{f_2})
\label{eq:pen_Proca}
\end{equation}
for general $f_i\in\Omega_0^1(\M)$. In the Maxwell case, we apply
Proposition~\ref{lem:eom_test}(b), axiom M3 and Eq.~(\ref{eq:locHad_em})
to obtain
\begin{equation}
\omega\left( [\A](f_1) [\A](f_2) \right) =
\omega\left( [\A] (\widetilde{f_1}) [\A] (\widetilde{f_2}) \right)
=W(\widetilde{f_1},\widetilde{f_2})
\label{eq:pen_em}
\end{equation}
for co-closed $f_i\in\Omega_0^1(\M)$. The following claim will be proved
below.

\begin{Proposition} \label{prop:global}
In the Proca case, 
\begin{equation}
W_\mass((-\bdelta\bd+\mass^2)f,f') =
W_\mass(f',(-\bdelta\bd+\mass^2)f)=0
\label{eq:glob_proc}
\end{equation}
for all $f,f'\in\Omega_0^1(\M)$, while in the Maxwell case, we have 
\begin{equation}
W(\bdelta\bd f,f') = W(f',\bdelta\bd f)=0
\label{eq:glob_em}
\end{equation}
for all $f,f'\in\Omega_0^1(\M)$ with $f'$ co-closed.
\end{Proposition}

In combination with the explicit form of
$\widetilde{f}_i-f_i$ (and, in the Proca case, the fact that 
$(\id-\mass^{-2}\bd\bdelta)$ and $-\bdelta\bd+\mass^2$ commute)
Prop.~\ref{prop:global} allows us
to show that
\begin{equation}
W_\mass(\widetilde{f_1},(\id-\mass^{-2}\bd\bdelta)\widetilde{f_2}) = 
W_\mass(f_1,(\id-\mass^{-2}\bd\bdelta)f_2)
\end{equation}
in the Proca case, and that $W(\widetilde{f_1},\widetilde{f_2}) =
W(f_1,f_2)$ holds for the Maxwell field. Taken together
with Eqs.~(\ref{eq:pen_Proca}) and~(\ref{eq:pen_em}) respectively, these
relations then establish Eqs.~(\ref{eq:locHad_Proca}) and~(\ref{eq:locHad_em}).
$\Box$

The claim made above is proved as follows.\\
{\noindent\em Proof of Proposition~\ref{prop:global}.} Fix an
arbitrary
$f'\in\Omega_0^1(\N)$ (with the additional requirement that $\bdelta
f'=0$ in the Maxwell case). Then $W_\mass(\cdot,f')$ and
$W_\mass(f',\cdot)$  (or $W(\cdot,f')$ and $W(f',\cdot)$ in 
the Maxwell case) obey the hypotheses of Lemma~\ref{lem:glob_lem}(a)
owing to Eq.~(\ref{eq:locHad_Proca}) and axiom P3 (resp., 
Eq.~(\ref{eq:locHad_em}) and axiom M3). Accordingly, 
Eq.~(\ref{eq:glob_proc}) (resp., Eq.~(\ref{eq:glob_em}))
holds for all $f\in\Omega_0^1(\M)$ and the fixed $f'\in\Omega_0^1(\N)$. 
Now fix $f\in\Omega_0^1(\M)$ arbitrarily. In the Proca case, 
$W_\mass((-\bdelta\bd+\mass^2)f,\cdot)$ and
$W_\mass(\cdot,(-\bdelta\bd+\mass^2)f)$ are weak
$(\Box+\mass^2)$-solutions vanishing on $\N$ and hence globally,
as required. In the Maxwell case, $W(\bdelta\bd f,\cdot)$ and
$W(\cdot,\bdelta\bd f)$ are weak $\Box$-solutions vanishing
on co-closed elements of $\Omega_0^1(\N)$. Lemma~\ref{lem:glob_lem}(b)
entails that they therefore vanish on all co-closed $f'\in\Omega_0^1(\M)$, thereby 
completing the proof.~$\Box$


\section{Construction of a Hadamard $(\Box+\mass^2)$-bisolution in ultrastatic
spacetimes}
\label{appx:bisolution}

In this appendix we prove Theorem~\ref{thm:Hadbisoln} by explicitly constructing
a Hadamard $(\Box+\mass^2)$-bisolution $W_\mass$ on any ultrastatic spacetime
$(\M,g)$ obeying the assumptions stated at the beginning of
Sect.~\ref{sect:ultra}. 
For $\mass>0$, the argument proceeds as follows.
First, we use functional calculus on the Hilbert space $\HH$ 
to define $W_\mass$ as a bilinear map from
$\Omega_0^1(\M)\times\Omega_0^1(\M)$ to $\complex$. We show that
$W_\mass$ is in fact a one-form bidistributional weak 
$(\Box+\mass^2)$-bisolution, with antisymmetric part $-iE_\mass$, and
determine a crude bound on its wave-front set. Next, we appeal to the
existence of a {\em Hadamard} $(\Box+\mass^2)$-bisolution $\widetilde{W}_\mass$, also
with antisymmetric part $-iE_\mass$, established in Lemma~5.4(a) 
of~\cite{Sah&Ver}. A simple microlocal argument is used to show that
$W_\mass-\widetilde{W}_\mass$ is smooth, from which it follows that
$W_\mass$ is itself Hadamard. Finally, we show that $W_\mass$ may be
expanded in terms of any $\KK$-pseudo-orthonormal complete set of
eigenvectors for the operator $K$ as claimed in Theorem~\ref{thm:Hadbisoln}. 
The argument is only slightly different in the case $\mass=0$. 

It will be convenient to regard each $\A\in\Omega^1(\M)$ as a smooth
one-parameter family $t\mapsto \A(t)$ of elements in
$\HH=\Lambda^0(\Sigma)\oplus\Lambda^1(\Sigma)$,
where $\A(t)$ is the restriction of $\A$ to the constant time surface
$\{t\}\times\Sigma$. The pairing $\langle\cdot,\cdot\rangle$ of
one-forms on $\M$ is related to the inner products of $\HH$ and $\KK$ by
\begin{equation}
\langle \A,\B\rangle = \int dt\, \dip{\overline{\A(t)}}{\B(t)} =
\int dt\, \left(\overline{\A(t)},J\B(t)\right)_\HH\,,
\label{eq:ipreln}
\end{equation}
where $J=\id_{\Lambda^0(\Sigma)}\oplus -\id_{\Lambda^1(\Sigma)}$. We
note that $J$ and $K=(-\Delta^s_\Sigma+\mass^2)\oplus(-\Delta_\Sigma+\mass^2)$
commute.

With these conventions, the Klein--Gordon equation $(\Box+\mass^2)\A=0$ may 
be written as the Hilbert space ordinary differential equation 
\begin{equation}
-\partial_t^2 \A(t) = K\A(t)\,.
\end{equation}
Suppose that $\mass>0$. Because $\Sigma$ is compact, $K$ has  
discrete spectrum bounded below by $\mass^2>0$. Accordingly, the operator
$K^{-1/2}$ is well-defined and bounded, and the 
advanced-minus-retarded solution operator $E_\mass:\Omega_0^1(\M)\to\Omega^1(\M)$
may be written
\begin{equation}
(E_\mass \J)(t) = \int dt'\, K^{-1/2}\sin \left[K^{1/2}(t'-t)\right] \J(t')\,,
\end{equation}
thus obtaining
\begin{equation}
E_\mass(\J,\J') = \langle \J,E_\mass \J'\rangle = 
\int dt\,dt'\, 
\left(\overline{\J(t)},JK^{-1/2}\sin \left[K^{1/2}(t'-t)\right] \J'(t')\right)_\HH\,.
\end{equation}

We define our candidate Hadamard $(\Box+\mass^2)$-bisolution $W_\mass$
by taking the positive frequency part of $-iE_\mass$, i.e., by replacing
the sine function by an exponential. To be precise,
for $f,f'\in\D(\real)$ and
$g,g'\in\Omega^0(\Sigma)\oplus\Omega^1(\Sigma)$, we define
\begin{equation}
W_\mass(f\otimes g,f'\otimes g')= -\frac{1}{2} 
\left(\widehat{\overline{f}}(K^{1/2})\overline{g}, K^{-1/2}
\widehat{f'}(K^{1/2})Jg'\right)_\HH\,,
\end{equation}
in which operators such as $\widehat{f'}(K^{1/2})$ are defined by 
functional calculus. Using the Cauchy--Schwarz inequality and elementary
operator norm estimates, we find
\begin{equation}
\left| W_\mass(f\otimes g,f'\otimes g')\right|\le \frac{1}{2} 
\|K^{-1/2}\|\, \|g\| \,\|g'\| \sup_{\omega\in\sigma(K^{1/2})} |\widehat{f}(-\omega)| 
\sup_{\omega\in\sigma(K^{1/2})} |\widehat{f'}(\omega)| \,,
\label{eq:explicit}
\end{equation}
from which it follows that $W_\mass$ extends to a distribution
in $(\Omega_0^1(\M)\times\Omega_0^1(\M))'$. It is 
straightforward to check that $W_\mass$ is a weak
$(\Box+\mass^2)$-bisolution, with antisymmetric part $-iE_\mass$.

The wave-front set of $W_\mass$ may be estimated in two ways. First,
because it is a $(\Box+\mass^2)$-bisolution, we have
\begin{equation}
\WF(W_\mass)\subseteq \N\times\N 
\end{equation}
where $\N =\{(x,k)\in T^*\M: g^{ab}(x)k_a k_b=0\}$ is the null
bundle of $(\M,g)$. Secondly, the explicit bound~(\ref{eq:explicit}),
coupled with the observation that 
$\sup_{\omega\in\sigma(K^{1/2})}|\widehat{f}(\lambda-\omega)|$ is
bounded as $\lambda\to+\infty$, but rapidly decaying for
$\lambda\to-\infty$ (and vice versa for 
$\sup_{\omega\in\sigma(K^{1/2})}|\widehat{f}(\lambda+\omega)|$) entails
that
\begin{equation}
\WF(W_\mass)\subseteq (\T^+\cup\Z)\times(\T^-\cup\Z)
\end{equation}
where $\T^\pm=\{(x,k)\in T^*\M: \pm k_0>0\}$ and $\Z=\{(x,k)\in T^*\M:
k=0\}$ is the zero section of $T^*\M$. Comparing these two estimates, 
\begin{equation}
\WF(W_\mass)\subseteq (\N^+\cup\Z)\times(\N^-\cup\Z)
\label{eq:wfbd}
\end{equation}
where $\N^\pm = \N\cap\T^\pm$ are the future ($+$) and past ($-$) 
null bundles. 

We now appeal to the existence of a Hadamard form
$(\Box+\mass^2)$-bisolution $\widetilde{W}_\mass$ on $(\M,g)$ with
antisymmetric part $-iE_\mass$ (Lemma~5.4(a) of~\cite{Sah&Ver}). 
Since the wave-front set of 
$\widetilde{W}_\mass$ also obeys~(\ref{eq:wfbd}), we have
\begin{equation}
\WF(W_\mass - \widetilde{W}_\mass)\subseteq (\N^+\cup\Z)\times(\N^-\cup\Z)\,.
\end{equation}
But $W_\mass - \widetilde{W}_\mass$ is symmetric, so we also have
\begin{equation}
\WF(W_\mass - \widetilde{W}_\mass)\subseteq (\N^-\cup\Z)\times(\N^+\cup\Z)\,.
\end{equation}
Comparing these two bounds, we see that $\WF(W_\mass -
\widetilde{W}_\mass)\subseteq
\Z\times\Z$ and, since the wave-front set excludes the zero section, we
conclude that this wave-front set is in fact empty. Accordingly, 
$W_\mass=\widetilde{W}_\mass$ (mod~$C^\infty$) so $W_\mass$ 
is of Hadamard form.

The analysis of the massless case is complicated by the
existence of a zero eigenvalue mode $\varphi_0=(1,0)$ for $K$.
(Triviality of $H_1(\Sigma)$ precludes the existence of any harmonic
one-forms on $\Sigma$, so $\varphi_0$ is the unique zero mode.)
However, the spectrum of $K$ is otherwise bounded away from zero,
so $K^{-1/2}$ is well-defined and bounded on $P\HH$, where
$P$ is the orthogonal projector onto the orthogonal complement of
$\varphi_0$.  
In this case, we define
\begin{equation}
W(f\otimes g,f'\otimes g')= -\frac{1}{2} 
\left(\widehat{\overline{f}}(K^{1/2})P\overline{g}, K^{-1/2}
\widehat{f'}(K^{1/2})PJg'\right)_\HH\,.
\end{equation}
It is easy to check that $W$ is  a bidistributional
$\Box$-bisolution, whose antisymmetric part is
\begin{equation}
W(\J,\J')-W(\J',\J) =-iE_0(\J,\J') +i
\int dt\,dt'\, (t'-t)\,(\overline{\J(t)},\,\varphi_0)_\HH\,
(\varphi_0,J\J'(t'))_\HH
\label{eq:B13}
\end{equation}
and is therefore equal (mod~$C^\infty$) to $-iE_0$. Appealing as before to
the existence of a Hadamard $\Box$-bisolution $\widetilde{W}$ with
antisymmetric part $-iE_0$, the argument used above shows
that $W=\widetilde{W}$ (mod~$C^\infty$) because $W-\widetilde{W}$ is
symmetric (mod~$C^\infty$). It remains to show that the last term
in Eq.~(\ref{eq:B13}) vanishes if $\J$ and $\J'$ are both co-closed, 
as required for consistency with the commutator axiom M4. Using
Eq.~(\ref{eq:ipreln}) and the fact that $J\varphi_0=\varphi_0$, one may show
that the term in question is proportional to
$\langle\overline{\J},\bd t\rangle\,\langle t\bd t,\J'\rangle -
\langle\overline{\J},t\bd t\rangle\,\langle \bd t,\J'\rangle$, which
vanishes because $\langle\overline{\J},\bd t\rangle=\langle
\overline{\bdelta\J},t\rangle=0$ and similarly 
$\langle \bd t,\J'\rangle=0$. 

Finally, let $\xi_j$ be a complete set of $\KK$-pseudo-orthonormal 
$K$-eigenfunctions, with corresponding eigenvalues $\omega_\lambda^2$
($\omega_\lambda\ge 0$). Using the completeness relation
Eq.~(\ref{eq:completeness}), we see that
\begin{equation}
\widehat{f'}(K^{1/2})^*K^{-1/2}\widehat{\overline{f}}(K^{1/2})P\overline{g}
=\sum_{j:\omega_j\not=0} \xi_j \omega_j^{-1}
\overline{\widehat{f'}(\omega_j)}\widehat{\overline{f}}(\omega_j)
\dip{\xi_j}{\xi_j}\,\dip{\xi_j}{\overline{g}}
\end{equation}
and hence
\begin{eqnarray}
W(\J,\J') &=& -\sum_{j:\omega_j\not=0}\frac{1}{2\omega_j}
\dip{\xi_j}{\xi_j}\,
\dip{\widehat{\overline{f}}(\omega_j)\overline{g}}{\xi_j}
\dip{\xi_j}{\widehat{f'}(\omega_j)g'} \nonumber\\
&=&-\sum_{j:\omega_j\not=0}\frac{1}{2\omega_j}
\dip{\xi_j}{\xi_j}\, \langle \A_j,\J\rangle\,\langle\overline{\A_j},\J'\rangle\,,
\label{eq:WJJ}
\end{eqnarray}
where we have defined the modes $\A_j$ by 
$\A_j(t)=e^{-i\omega_j t} \xi_\lambda$ and used
\begin{equation}
\dip{\xi_j}{\widehat{f'}(\omega_j)g'} = \int dt\, e^{i\omega_j t}\dip{\xi_j}{f'(t)g'}
= \int dt\, \dip{\A_j(t)}{\J'(t)}
= \langle \overline{\A_j},\J'\rangle
\end{equation}
and $\dip{\widehat{\overline{f}}(\omega_j)\overline{g}}{\xi_j}=\langle
\J,\A_j\rangle =\langle \A_j,\J\rangle$. 

In the case $\mass>0$, exactly analogous arguments show that the
$W_\mass(\J,\J')$ may also be written in the form of the right-hand side
of Eq.~(\ref{eq:WJJ}) [the restriction to $\omega_j>0$ is
inessential as all modes obey this condition]. This completes the
proof of Theorem~\ref{thm:Hadbisoln}.

\begin{acknowledgments}
Part of this work was conducted at 
the Erwin Schr\"{o}dinger Institute for Mathematical Physics,
Vienna, during the programme 
on Quantum Field Theory in Curved Spacetime. 
We are grateful to the organisers of this programme and to the
Institute for its hospitality and financial support. We also thank Atsushi Higuchi,
Wolfgang Junker, Fernando Lled\'{o}, Ian McIntosh and Rainer Verch 
for many illuminating discussions.
This work was partially supported by EPSRC Grant GR/R25019/01 to the
University of York. MJP also thanks the University of York
for a grant awarded under its ``Research Funding for Staff on Fixed Term
Contracts'' scheme.
\end{acknowledgments}

\begin{thebibliography}{65}
\expandafter\ifx\csname natexlab\endcsname\relax\def\natexlab#1{#1}\fi
\expandafter\ifx\csname bibnamefont\endcsname\relax
  \def\bibnamefont#1{#1}\fi
\expandafter\ifx\csname bibfnamefont\endcsname\relax
  \def\bibfnamefont#1{#1}\fi
\expandafter\ifx\csname citenamefont\endcsname\relax
  \def\citenamefont#1{#1}\fi
\expandafter\ifx\csname url\endcsname\relax
  \def\url#1{\texttt{#1}}\fi
\expandafter\ifx\csname urlprefix\endcsname\relax\def\urlprefix{URL }\fi
\providecommand{\bibinfo}[2]{#2}
\providecommand{\eprint}[2][]{\url{#2}}

\bibitem[{\citenamefont{Penrose}(1965)}]{Penrose}
\bibinfo{author}{\bibfnamefont{R.}~\bibnamefont{Penrose}},
  \bibinfo{journal}{Phys. Rev. Lett.} \textbf{\bibinfo{volume}{14}},
  \bibinfo{pages}{57} (\bibinfo{year}{1965}).

\bibitem[{\citenamefont{Hawking}(1965)}]{Hawking}
\bibinfo{author}{\bibfnamefont{S.~W.} \bibnamefont{Hawking}},
  \bibinfo{journal}{Phys. Rev. Lett.} \textbf{\bibinfo{volume}{15}},
  \bibinfo{pages}{689} (\bibinfo{year}{1965}).

\bibitem[{\citenamefont{Epstein et~al.}(1965)\citenamefont{Epstein, Glaser, and
  Jaffe}}]{Epstein}
\bibinfo{author}{\bibfnamefont{H.}~\bibnamefont{Epstein}},
  \bibinfo{author}{\bibfnamefont{V.}~\bibnamefont{Glaser}}, \bibnamefont{and}
  \bibinfo{author}{\bibfnamefont{A.}~\bibnamefont{Jaffe}}, \bibinfo{journal}{Il
  Nuovo Cim.} \textbf{\bibinfo{volume}{36}}, \bibinfo{pages}{1016}
  (\bibinfo{year}{1965}).

\bibitem[{\citenamefont{Morris and Thorne}(1988)}]{Morris88a}
\bibinfo{author}{\bibfnamefont{M.~S.} \bibnamefont{Morris}} \bibnamefont{and}
  \bibinfo{author}{\bibfnamefont{K.~S.} \bibnamefont{Thorne}},
  \bibinfo{journal}{Am. J. Phys.} \textbf{\bibinfo{volume}{56}},
  \bibinfo{pages}{395} (\bibinfo{year}{1988}).

\bibitem[{\citenamefont{Morris et~al.}(1988)\citenamefont{Morris, Thorne, and
  Yurtsever}}]{Morris88b}
\bibinfo{author}{\bibfnamefont{M.~S.} \bibnamefont{Morris}},
  \bibinfo{author}{\bibfnamefont{K.~S.} \bibnamefont{Thorne}},
  \bibnamefont{and}
  \bibinfo{author}{\bibfnamefont{U.}~\bibnamefont{Yurtsever}},
  \bibinfo{journal}{Phys. Rev. Lett.} \textbf{\bibinfo{volume}{61}},
  \bibinfo{pages}{1446} (\bibinfo{year}{1988}).

\bibitem[{\citenamefont{Hiscock}(1981)}]{Hisc81}
\bibinfo{author}{\bibfnamefont{W.~A.} \bibnamefont{Hiscock}},
  \bibinfo{journal}{Ann. Phys.} \textbf{\bibinfo{volume}{131}},
  \bibinfo{pages}{245} (\bibinfo{year}{1981}).

\bibitem[{\citenamefont{Ford and Roman}(1992)}]{F&Ro92}
\bibinfo{author}{\bibfnamefont{L.~H.} \bibnamefont{Ford}} \bibnamefont{and}
  \bibinfo{author}{\bibfnamefont{T.~A.} \bibnamefont{Roman}},
  \bibinfo{journal}{Phys. Rev. D} \textbf{\bibinfo{volume}{46}},
  \bibinfo{pages}{1328} (\bibinfo{year}{1992}).

\bibitem[{\citenamefont{Alcubierre}(1994)}]{Alcu94}
\bibinfo{author}{\bibfnamefont{M.}~\bibnamefont{Alcubierre}},
  \bibinfo{journal}{Class. Quantum Grav.} \textbf{\bibinfo{volume}{11}},
  \bibinfo{pages}{L73} (\bibinfo{year}{1994}).

\bibitem[{\citenamefont{Krasnikov}(1998)}]{Kras95}
\bibinfo{author}{\bibfnamefont{S.~V.} \bibnamefont{Krasnikov}},
  \bibinfo{journal}{Phys. Rev. D} \textbf{\bibinfo{volume}{57}},
  \bibinfo{pages}{4760} (\bibinfo{year}{1998}), \bibinfo{note}{gr-qc/9511068}.

\bibitem[{\citenamefont{Everett}(1996)}]{Ever94}
\bibinfo{author}{\bibfnamefont{A.~E.} \bibnamefont{Everett}},
  \bibinfo{journal}{Phys. Rev. D} \textbf{\bibinfo{volume}{53}},
  \bibinfo{pages}{7365} (\bibinfo{year}{1996}).

\bibitem[{\citenamefont{Parker and Fulling}(1973)}]{P&Fu73}
\bibinfo{author}{\bibfnamefont{L.}~\bibnamefont{Parker}} \bibnamefont{and}
  \bibinfo{author}{\bibfnamefont{S.~A.} \bibnamefont{Fulling}},
  \bibinfo{journal}{Phys. Rev. D} \textbf{\bibinfo{volume}{7}},
  \bibinfo{pages}{2357} (\bibinfo{year}{1973}).

\bibitem[{\citenamefont{Ford}(1978)}]{Ford78}
\bibinfo{author}{\bibfnamefont{L.~H.} \bibnamefont{Ford}},
  \bibinfo{journal}{Proc. Roy. Soc. Lond. A} \textbf{\bibinfo{volume}{364}},
  \bibinfo{pages}{227} (\bibinfo{year}{1978}).

\bibitem[{\citenamefont{Ford}(1991)}]{Ford91}
\bibinfo{author}{\bibfnamefont{L.~H.} \bibnamefont{Ford}},
  \bibinfo{journal}{Phys. Rev. D} \textbf{\bibinfo{volume}{43}},
  \bibinfo{pages}{3972} (\bibinfo{year}{1991}).

\bibitem[{\citenamefont{Ford and Roman}(1995)}]{F&Ro95}
\bibinfo{author}{\bibfnamefont{L.~H.} \bibnamefont{Ford}} \bibnamefont{and}
  \bibinfo{author}{\bibfnamefont{T.~A.} \bibnamefont{Roman}},
  \bibinfo{journal}{Phys. Rev. D} \textbf{\bibinfo{volume}{51}},
  \bibinfo{pages}{4277} (\bibinfo{year}{1995}), \bibinfo{note}{gr-qc/9410043}.

\bibitem[{\citenamefont{Ford and Roman}(1997)}]{F&Ro97}
\bibinfo{author}{\bibfnamefont{L.~H.} \bibnamefont{Ford}} \bibnamefont{and}
  \bibinfo{author}{\bibfnamefont{T.~A.} \bibnamefont{Roman}},
  \bibinfo{journal}{Phys. Rev. D} \textbf{\bibinfo{volume}{55}},
  \bibinfo{pages}{2082} (\bibinfo{year}{1997}), \bibinfo{note}{gr-qc/9607003}.

\bibitem[{\citenamefont{Pfenning and Ford}(1997)}]{Pfen97a}
\bibinfo{author}{\bibfnamefont{M.~J.} \bibnamefont{Pfenning}} \bibnamefont{and}
  \bibinfo{author}{\bibfnamefont{L.~H.} \bibnamefont{Ford}},
  \bibinfo{journal}{Phys. Rev. D} \textbf{\bibinfo{volume}{55}},
  \bibinfo{pages}{4813} (\bibinfo{year}{1997}), \bibinfo{note}{gr-qc/9608005}.

\bibitem[{\citenamefont{Pfenning and Ford}(1998)}]{Pfen98a}
\bibinfo{author}{\bibfnamefont{M.~J.} \bibnamefont{Pfenning}} \bibnamefont{and}
  \bibinfo{author}{\bibfnamefont{L.~H.} \bibnamefont{Ford}},
  \bibinfo{journal}{Phys. Rev. D} \textbf{\bibinfo{volume}{57}},
  \bibinfo{pages}{3489} (\bibinfo{year}{1998}), \bibinfo{note}{gr-qc/9710055}.

\bibitem[{\citenamefont{Pfenning}(1998)}]{Pfen98b}
\bibinfo{author}{\bibfnamefont{M.~J.} \bibnamefont{Pfenning}}, Ph.D. thesis,
  \bibinfo{school}{Tufts University}, \bibinfo{address}{Medford, Massachusetts}
  (\bibinfo{year}{1998}), \bibinfo{note}{gr-qc/9805037}.

\bibitem[{\citenamefont{Flanagan}(1997)}]{Flan97}
\bibinfo{author}{\bibfnamefont{{\'{E}}.~{\'{E}}.} \bibnamefont{Flanagan}},
  \bibinfo{journal}{Phys. Rev. D} \textbf{\bibinfo{volume}{56}},
  \bibinfo{pages}{4922} (\bibinfo{year}{1997}), \bibinfo{note}{gr-qc/9706006}.

\bibitem[{\citenamefont{Flanagan}(2002)}]{Flan02}
\bibinfo{author}{\bibfnamefont{{\'{E}}.~{\'{E}}.} \bibnamefont{Flanagan}},
  \bibinfo{journal}{Phys. Rev. D} \textbf{\bibinfo{volume}{66}},
  \bibinfo{pages}{104007} (\bibinfo{year}{2002}),
  \bibinfo{note}{gr-qc/0208066}.

\bibitem[{\citenamefont{Vollick}(2000)}]{Voll00}
\bibinfo{author}{\bibfnamefont{D.~N.} \bibnamefont{Vollick}},
  \bibinfo{journal}{Phys. Rev. D} \textbf{\bibinfo{volume}{61}},
  \bibinfo{pages}{084022} (\bibinfo{year}{2000}).

\bibitem[{\citenamefont{Fewster and Eveson}(1998)}]{Fe&E98}
\bibinfo{author}{\bibfnamefont{C.~J.} \bibnamefont{Fewster}} \bibnamefont{and}
  \bibinfo{author}{\bibfnamefont{S.~P.} \bibnamefont{Eveson}},
  \bibinfo{journal}{Phys. Rev. D} \textbf{\bibinfo{volume}{58}},
  \bibinfo{pages}{084010} (\bibinfo{year}{1998}),
  \bibinfo{note}{gr-qc/9805024}.

\bibitem[{\citenamefont{Fewster and Teo}(1999)}]{Fe&T99}
\bibinfo{author}{\bibfnamefont{C.~J.} \bibnamefont{Fewster}} \bibnamefont{and}
  \bibinfo{author}{\bibfnamefont{E.}~\bibnamefont{Teo}},
  \bibinfo{journal}{Phys. Rev. D} \textbf{\bibinfo{volume}{59}},
  \bibinfo{pages}{104016} (\bibinfo{year}{1999}).

\bibitem[{\citenamefont{Fewster}(2000)}]{Fews00}
\bibinfo{author}{\bibfnamefont{C.~J.} \bibnamefont{Fewster}},
  \bibinfo{journal}{Class. Quantum Grav.} \textbf{\bibinfo{volume}{17}},
  \bibinfo{pages}{1897} (\bibinfo{year}{2000}).

\bibitem[{\citenamefont{Fewster and Verch}(2002)}]{Fe&V02}
\bibinfo{author}{\bibfnamefont{C.~J.} \bibnamefont{Fewster}} \bibnamefont{and}
  \bibinfo{author}{\bibfnamefont{R.}~\bibnamefont{Verch}},
  \bibinfo{journal}{Commun. Math. Phys.} \textbf{\bibinfo{volume}{225}},
  \bibinfo{pages}{331} (\bibinfo{year}{2002}).

\bibitem[{\citenamefont{Helfer}()}]{HelferII}
\bibinfo{author}{\bibfnamefont{A.~D.} \bibnamefont{Helfer}},
  \bibinfo{note}{hep-th/9908012}.

\bibitem[{\citenamefont{Pfenning}(2002)}]{Pfen02}
\bibinfo{author}{\bibfnamefont{M.~J.} \bibnamefont{Pfenning}},
  \bibinfo{journal}{Phys. Rev. D} \textbf{\bibinfo{volume}{65}},
  \bibinfo{pages}{024009} (\bibinfo{year}{2002}).

\bibitem[{\citenamefont{Marecki}(2002)}]{Marecki}
\bibinfo{author}{\bibfnamefont{P.}~\bibnamefont{Marecki}},
  \bibinfo{journal}{Phys. Rev. A} \textbf{\bibinfo{volume}{66}},
  \bibinfo{pages}{053801} (\bibinfo{year}{2002}),
  \bibinfo{note}{quant-ph/0203027}.

\bibitem[{\citenamefont{Radzikowski}(1996)}]{Radz96}
\bibinfo{author}{\bibfnamefont{M.~J.} \bibnamefont{Radzikowski}},
  \bibinfo{journal}{Commun. Math. Phys.} \textbf{\bibinfo{volume}{179}},
  \bibinfo{pages}{529} (\bibinfo{year}{1996}).

\bibitem[{\citenamefont{Kratzert}(2000)}]{Kratzert}
\bibinfo{author}{\bibfnamefont{K.}~\bibnamefont{Kratzert}},
  \bibinfo{journal}{Annalen Phys.} \textbf{\bibinfo{volume}{9}},
  \bibinfo{pages}{475} (\bibinfo{year}{2000}).

\bibitem[{\citenamefont{Hollands}(2001)}]{Hollands}
\bibinfo{author}{\bibfnamefont{S.}~\bibnamefont{Hollands}},
  \bibinfo{journal}{Commun. Math. Phys.} \textbf{\bibinfo{volume}{216}},
  \bibinfo{pages}{635} (\bibinfo{year}{2001}).

\bibitem[{\citenamefont{Sahlmann and Verch}(2001)}]{Sah&Ver}
\bibinfo{author}{\bibfnamefont{H.}~\bibnamefont{Sahlmann}} \bibnamefont{and}
  \bibinfo{author}{\bibfnamefont{R.}~\bibnamefont{Verch}},
  \bibinfo{journal}{Rev. Math. Phys.} \textbf{\bibinfo{volume}{13}},
  \bibinfo{pages}{1203} (\bibinfo{year}{2001}),
  \bibinfo{note}{math-ph/0008029}.

\bibitem[{\citenamefont{Brown and Ottewill}(1986)}]{B&OT86}
\bibinfo{author}{\bibfnamefont{M.~R.} \bibnamefont{Brown}} \bibnamefont{and}
  \bibinfo{author}{\bibfnamefont{A.~C.} \bibnamefont{Ottewill}},
  \bibinfo{journal}{Phys. Rev. D} \textbf{\bibinfo{volume}{34}},
  \bibinfo{pages}{1776} (\bibinfo{year}{1986}).

\bibitem[{\citenamefont{Allen and Ottewill}(1992)}]{AllOtt92}
\bibinfo{author}{\bibfnamefont{B.}~\bibnamefont{Allen}} \bibnamefont{and}
  \bibinfo{author}{\bibfnamefont{A.~C.} \bibnamefont{Ottewill}},
  \bibinfo{journal}{Phys. Rev. D} \textbf{\bibinfo{volume}{46}},
  \bibinfo{pages}{861} (\bibinfo{year}{1992}).

\bibitem[{\citenamefont{Junker and {Lled\'o}}()}]{JunLle}
\bibinfo{author}{\bibfnamefont{W.}~\bibnamefont{Junker}} \bibnamefont{and}
  \bibinfo{author}{\bibfnamefont{F.}~\bibnamefont{{Lled\'o}}},
  \bibinfo{note}{in preparation}.

\bibitem[{\citenamefont{Dimock}(1992)}]{Dimock}
\bibinfo{author}{\bibfnamefont{J.}~\bibnamefont{Dimock}},
  \bibinfo{journal}{Rev. Math. Phys.} \textbf{\bibinfo{volume}{4}},
  \bibinfo{pages}{223} (\bibinfo{year}{1992}).

\bibitem[{\citenamefont{Furlani}(1999)}]{Furl99}
\bibinfo{author}{\bibfnamefont{E.~P.} \bibnamefont{Furlani}},
  \bibinfo{journal}{J. Math. Phys.} \textbf{\bibinfo{volume}{40}},
  \bibinfo{pages}{2611} (\bibinfo{year}{1999}).

\bibitem[{\citenamefont{Fulling et~al.}(1981)\citenamefont{Fulling, Narcowich,
  and Wald}}]{FNW81}
\bibinfo{author}{\bibfnamefont{S.~A.} \bibnamefont{Fulling}},
  \bibinfo{author}{\bibfnamefont{F.~J.} \bibnamefont{Narcowich}},
  \bibnamefont{and} \bibinfo{author}{\bibfnamefont{R.~M.} \bibnamefont{Wald}},
  \bibinfo{journal}{Ann. Phys.} \textbf{\bibinfo{volume}{136}},
  \bibinfo{pages}{243} (\bibinfo{year}{1981}).

\bibitem[{\citenamefont{Abraham et~al.}(1988)\citenamefont{Abraham, Marsden,
  and Ratiu}}]{AMR88}
\bibinfo{author}{\bibfnamefont{R.}~\bibnamefont{Abraham}},
  \bibinfo{author}{\bibfnamefont{J.~E.} \bibnamefont{Marsden}},
  \bibnamefont{and} \bibinfo{author}{\bibfnamefont{T.}~\bibnamefont{Ratiu}},
  \emph{\bibinfo{title}{Manifolds, Tensor Analysis, and Applications}}
  (\bibinfo{publisher}{Springer-Verlag}, \bibinfo{address}{New York},
  \bibinfo{year}{1988}), \bibinfo{edition}{2nd} ed.

\bibitem[{\citenamefont{{H\"{o}rmander}}(1983)}]{Hormander_1}
\bibinfo{author}{\bibfnamefont{L.}~\bibnamefont{{H\"{o}rmander}}},
  \emph{\bibinfo{title}{The Analysis of Linear Partial Differential
  Operators}}, vol.~\bibinfo{volume}{I} (\bibinfo{publisher}{Springer-Verlag},
  \bibinfo{address}{Berlin}, \bibinfo{year}{1983}), \bibinfo{edition}{2nd} ed.

\bibitem[{\citenamefont{Dieckmann}(1988)}]{Dieck88}
\bibinfo{author}{\bibfnamefont{J.}~\bibnamefont{Dieckmann}},
  \bibinfo{journal}{J. Math. Phys.} \textbf{\bibinfo{volume}{29}},
  \bibinfo{pages}{578} (\bibinfo{year}{1988}).

\bibitem[{\citenamefont{{Choquet-Bruhat}}(1968)}]{Choq67}
\bibinfo{author}{\bibfnamefont{Y.}~\bibnamefont{{Choquet-Bruhat}}},
  \emph{\bibinfo{title}{Batelle Rencontres}} (\bibinfo{publisher}{Benjamin},
  \bibinfo{address}{New York}, \bibinfo{year}{1968}),
  chap.~\bibinfo{chapter}{IV}, pp. \bibinfo{pages}{{84--106}}.

\bibitem[{\citenamefont{Strohmaier}(2000)}]{Stro00}
\bibinfo{author}{\bibfnamefont{A.}~\bibnamefont{Strohmaier}},
  \bibinfo{journal}{Commun. Math. Phys.} \textbf{\bibinfo{volume}{215}},
  \bibinfo{pages}{105} (\bibinfo{year}{2000}).

\bibitem[{\citenamefont{Strocchi}(1967)}]{StrocchiI}
\bibinfo{author}{\bibfnamefont{F.}~\bibnamefont{Strocchi}},
  \bibinfo{journal}{Phys. Rev.} \textbf{\bibinfo{volume}{165}},
  \bibinfo{pages}{1429} (\bibinfo{year}{1967}).

\bibitem[{\citenamefont{Strocchi}(1970)}]{StrocchiIII}
\bibinfo{author}{\bibfnamefont{F.}~\bibnamefont{Strocchi}},
  \bibinfo{journal}{Phys. Rev. D} \textbf{\bibinfo{volume}{2}},
  \bibinfo{pages}{2334} (\bibinfo{year}{1970}).

\bibitem[{\citenamefont{Baez et~al.}(1992)\citenamefont{Baez, Segal, and
  Zhou}}]{BSZ}
\bibinfo{author}{\bibfnamefont{J.}~\bibnamefont{Baez}},
  \bibinfo{author}{\bibfnamefont{I.}~\bibnamefont{Segal}}, \bibnamefont{and}
  \bibinfo{author}{\bibfnamefont{Z.}~\bibnamefont{Zhou}},
  \emph{\bibinfo{title}{Introduction to Algebraic and Constructive Quantum
  Field Theory}} (\bibinfo{publisher}{Princeton University Press},
  \bibinfo{address}{Princeton}, \bibinfo{year}{1992}).

\bibitem[{\citenamefont{Kay and Wald}(1991)}]{Ka&Wa}
\bibinfo{author}{\bibfnamefont{B.~S.} \bibnamefont{Kay}} \bibnamefont{and}
  \bibinfo{author}{\bibfnamefont{R.~M.} \bibnamefont{Wald}},
  \bibinfo{journal}{Phys. Rep.} \textbf{\bibinfo{volume}{207}},
  \bibinfo{pages}{49} (\bibinfo{year}{1991}).

\bibitem[{\citenamefont{Lled\'o}(2000)}]{Fernando}
\bibinfo{author}{\bibfnamefont{F.}~\bibnamefont{Lled\'o}},
  \bibinfo{howpublished}{{talk at the meeting on Microlocal Analysis and
  Quantum Field Theory, Mathematisches Forschungsinstitut Oberwolfach}}
  (\bibinfo{year}{2000}).

\bibitem[{\citenamefont{Cycon et~al.}(1987)\citenamefont{Cycon, Froese, Kirsch,
  and Simon}}]{CFKS}
\bibinfo{author}{\bibfnamefont{H.~L.} \bibnamefont{Cycon}},
  \bibinfo{author}{\bibfnamefont{R.~G.} \bibnamefont{Froese}},
  \bibinfo{author}{\bibfnamefont{W.}~\bibnamefont{Kirsch}}, \bibnamefont{and}
  \bibinfo{author}{\bibfnamefont{B.}~\bibnamefont{Simon}},
  \emph{\bibinfo{title}{{Schr\"odinger} Operators}}
  (\bibinfo{publisher}{Springer-Verlag}, \bibinfo{address}{Berlin},
  \bibinfo{year}{1987}).

\bibitem[{\citenamefont{Gupta}(1950)}]{Gupta50}
\bibinfo{author}{\bibfnamefont{S.~N.} \bibnamefont{Gupta}},
  \bibinfo{journal}{Proc. Phys. Soc. London} \textbf{\bibinfo{volume}{A63}},
  \bibinfo{pages}{681} (\bibinfo{year}{1950}).

\bibitem[{\citenamefont{Gupta}(1977)}]{Gupta}
\bibinfo{author}{\bibfnamefont{S.~N.} \bibnamefont{Gupta}},
  \emph{\bibinfo{title}{Quantum Electrodynamics}} (\bibinfo{publisher}{Gordon
  and Breach Science Publishers, Inc.}, \bibinfo{address}{New York},
  \bibinfo{year}{1977}).

\bibitem[{\citenamefont{H{\"{o}}rmander}(1971)}]{Hormander_FIO1}
\bibinfo{author}{\bibfnamefont{L.}~\bibnamefont{H{\"{o}}rmander}},
  \bibinfo{journal}{Acta Math.} \textbf{\bibinfo{volume}{127}},
  \bibinfo{pages}{79} (\bibinfo{year}{1971}).

\bibitem[{\citenamefont{Grundling and Lled\'{o}}(2000)}]{Grundling_Lledo2000}
\bibinfo{author}{\bibfnamefont{H.}~\bibnamefont{Grundling}} \bibnamefont{and}
  \bibinfo{author}{\bibfnamefont{F.}~\bibnamefont{Lled\'{o}}},
  \bibinfo{journal}{Rev. Math. Phys.} \textbf{\bibinfo{volume}{12}},
  \bibinfo{pages}{1159} (\bibinfo{year}{2000}).

\bibitem[{\citenamefont{Itzykson and Zuber}(1980)}]{Itz&Zu}
\bibinfo{author}{\bibfnamefont{C.}~\bibnamefont{Itzykson}} \bibnamefont{and}
  \bibinfo{author}{\bibfnamefont{J.-B.} \bibnamefont{Zuber}},
  \emph{\bibinfo{title}{Quantum Field Theory}}
  (\bibinfo{publisher}{McGraw-Hill, Inc.}, \bibinfo{address}{New York},
  \bibinfo{year}{1980}).

\bibitem[{\citenamefont{Fewster and Verch}()}]{Fe&V03}
\bibinfo{author}{\bibfnamefont{C.~J.} \bibnamefont{Fewster}} \bibnamefont{and}
  \bibinfo{author}{\bibfnamefont{R.}~\bibnamefont{Verch}}, \bibinfo{note}{to
  appear in Commun. Math. Phys., math-ph/0203010}.

\bibitem[{\citenamefont{Lifshitz}(1946)}]{Lifs46}
\bibinfo{author}{\bibfnamefont{E.~M.} \bibnamefont{Lifshitz}},
  \bibinfo{journal}{J. Phys. (Moscow)} \textbf{\bibinfo{volume}{10}},
  \bibinfo{pages}{116} (\bibinfo{year}{1946}).

\bibitem[{\citenamefont{Parker}(1972)}]{Park72}
\bibinfo{author}{\bibfnamefont{L.}~\bibnamefont{Parker}},
  \bibinfo{journal}{Phys. Rev. D} \textbf{\bibinfo{volume}{5}},
  \bibinfo{pages}{2905} (\bibinfo{year}{1972}).

\bibitem[{\citenamefont{Mashhoon}(1973)}]{Mash73}
\bibinfo{author}{\bibfnamefont{B.}~\bibnamefont{Mashhoon}},
  \bibinfo{journal}{Phys. Rev. D} \textbf{\bibinfo{volume}{8}},
  \bibinfo{pages}{4297} (\bibinfo{year}{1973}).

\bibitem[{\citenamefont{Parker and Fulling}(1974)}]{Parker}
\bibinfo{author}{\bibfnamefont{L.}~\bibnamefont{Parker}} \bibnamefont{and}
  \bibinfo{author}{\bibfnamefont{S.~A.} \bibnamefont{Fulling}},
  \bibinfo{journal}{Phys. Rev. D} \textbf{\bibinfo{volume}{9}},
  \bibinfo{pages}{341} (\bibinfo{year}{1974}).

\bibitem[{\citenamefont{Ford}(1976)}]{Ford76}
\bibinfo{author}{\bibfnamefont{L.~H.} \bibnamefont{Ford}},
  \bibinfo{journal}{Phys. Rev. D} \textbf{\bibinfo{volume}{14}},
  \bibinfo{pages}{3304} (\bibinfo{year}{1976}).

\bibitem[{\citenamefont{Jackson}(1962)}]{Jackson}
\bibinfo{author}{\bibfnamefont{J.~D.} \bibnamefont{Jackson}},
  \emph{\bibinfo{title}{Classical Electrodynamics}} (\bibinfo{publisher}{John
  Wiley \& Sons}, \bibinfo{address}{New York}, \bibinfo{year}{1962}),
  \bibinfo{edition}{2nd} ed., \bibinfo{note}{see Section 7.5}.

\bibitem[{\citenamefont{Gradshteyn and Ryzhik}(1994)}]{Gradshteyn}
\bibinfo{author}{\bibfnamefont{I.~S.} \bibnamefont{Gradshteyn}}
  \bibnamefont{and} \bibinfo{author}{\bibfnamefont{I.~M.}
  \bibnamefont{Ryzhik}}, \emph{\bibinfo{title}{Table of Integrals, Series, and
  Products}} (\bibinfo{publisher}{Academic Press}, \bibinfo{address}{San
  Diego}, \bibinfo{year}{1994}), \bibinfo{edition}{5th} ed.

\bibitem[{\citenamefont{Bott and Tu}(1982)}]{BottTu}
\bibinfo{author}{\bibfnamefont{R.}~\bibnamefont{Bott}} \bibnamefont{and}
  \bibinfo{author}{\bibfnamefont{L.~W.} \bibnamefont{Tu}},
  \emph{\bibinfo{title}{Differential forms in algebraic topology}}
  (\bibinfo{publisher}{Springer}, \bibinfo{address}{New York},
  \bibinfo{year}{1982}).

\bibitem[{\citenamefont{Warner}(1971)}]{Warner71}
\bibinfo{author}{\bibfnamefont{F.~W.} \bibnamefont{Warner}},
  \emph{\bibinfo{title}{Foundations of Differentiable Manifolds and Lie
  Groups}} (\bibinfo{publisher}{Scott, Foresman and Company},
  \bibinfo{address}{Glenview}, \bibinfo{year}{1971}).

\bibitem[{\citenamefont{Adams}(1975)}]{Adams75}
\bibinfo{author}{\bibfnamefont{R.}~\bibnamefont{Adams}},
  \emph{\bibinfo{title}{Sobolev Spaces}} (\bibinfo{publisher}{Academic Press},
  \bibinfo{address}{New York}, \bibinfo{year}{1975}).

\end{thebibliography}

\newpage

\end{document}